%% file: cluster-lensing-rar.tex
\documentclass[numberedappendix,twocolumn]{openjournal}
\usepackage[utf8]{inputenc}
\usepackage{amsmath}
\usepackage[dvipsnames]{xcolor}
\usepackage[colorlinks,allcolors=Blue]{hyperref}

\newcommand{\oneh}{\mathrm{1h}}
\newcommand{\twoh}{\mathrm{2h}}

\begin{document}

\title{Mass models of galaxy clusters from a non-parametric weak-lensing reconstruction}

\author{Tobias Mistele$^1$, Federico Lelli$^2$, Stacy McGaugh$^1$, James Schombert$^3$, Benoit Famaey$^4$}
\affiliation{%
$^1$Department of Astronomy, Case Western Reserve University, 10900 Euclid Avenue, Cleveland, Ohio 44106, USA \\
$^2$INAF -- Arcetri Astrophysical Observatory, Largo Enrico Fermi 5, 50125 Firenze, Italy \\
$^3$Department of Physics, University of Oregon, 1371 E 13th Ave, Eugene, Oregon 97403, USA \\
$^4$Université de Strasbourg, CNRS, Observatoire astronomique de Strasbourg, UMR 7550, F-67000 Strasbourg, France
}

\begin{abstract}
We study the CLASH sample of galaxy clusters using a new deprojection method for weak gravitational lensing observations.
This method is non-parametric, allowing us to infer mass profiles, or equivalently circular velocities, without having to assume a specific halo profile.
While this method assumes spherical symmetry,  we show that, on average, triaxiality is unlikely to significantly affect our results.
We use this method to study the total mass profiles of the CLASH clusters, as well as the relation between their total and baryonic components:
(1) We find that the implied circular velocities are consistent with being approximately flat at large radii, akin to the rotation curves of galaxies.
(2) We infer radially resolved baryonic mass fractions, finding that these vary significantly from cluster to cluster and depend strongly on the details of the X-ray gas mass profiles.
Since the gas mass profiles are poorly constrained at large radii, it is unclear whether the CLASH clusters reach the cosmic baryon fraction expected in $\Lambda$CDM.
(3) The non-parametric masses are consistent with the stellar mass--halo mass relation expected in $\Lambda$CDM.
(4) Galaxy clusters systematically deviate from the Baryonic Tully-Fisher Relation (BTFR) and the Radial Acceleration Relation (RAR) defined by galaxies, but the magnitude of the offset depends strongly on the gas mass extrapolation at large radii.
Contrary to some previous results based on hydrostatic equilibrium, we find that galaxy clusters may fall on the same BTFR and RAR as galaxies if one adds a suitable positive baryonic mass component.
\end{abstract}

\section{Introduction} 

Galaxy clusters are important astrophysical and cosmological probes.
For example, in cosmological models such as $\Lambda$CDM, their abundance and masses constrain key parameters such as $\Omega_m$ and $\sigma_8$.
They can also constrain models of dark matter and modified gravity, for example through features in their density profiles such as the splashback radius \citep[e.g.][]{Diemer2014,More2015b,More2016,Adhikari2018} or through scaling relations that connect their baryonic and dynamical mass distributions \citep[e.g.][]{Sanders2003,Tian2020b,Eckert2022,Li2023,Li2024b,Famaey2024,Kelleher2024}.

Scaling relations also play an important role in galaxies.
Indeed, galaxies follow tight scaling relations such as the Baryonic Tully-Fisher Relation \citep[BTFR,][]{McGaugh2000,Mistele2024} and the Radial Acceleration Relation \citep[RAR,][]{Lelli2017b,Brouwer2021,Mistele2023d}.
These were predicted a priori by Modified Newtonian Dynamics (MOND, \citealp{Milgrom1983a,Milgrom1983b,Milgrom1983c}, see \citealp{Famaey2025} for a recent review), but are not easily explained in $\Lambda$CDM because they must emerge from the complex and stochastic process of galaxy formation.
However, MOND-inspired theories have historically struggled to explain why galaxy clusters do not seem to follow the same scaling relations as galaxies \citep{Sanders1999,Sanders2003}.
Relativistic extension of MOND such as Aether-Scalar Tensor Theory \citep[AeST,][]{Skordis2020} or Relativistic Khronon Theory \citep{Blanchet2024} may ameliorate these issues, since they predict deviations from scaling relations such as the RAR at large masses \citep{Mistele2023,Durakovic2023}, but this has not yet been demonstrated to work in quantitative detail.

All these different constraints require dynamical masses of galaxy clusters.
A powerful tool to measure these is weak gravitational lensing.
Indeed, unlike X-ray and galaxy kinematics measurements, weak lensing does not require hydrostatic or dynamical equilibrium.
However, most existing weak-lensing measurements are based on fitting weak-lensing observations to parametric profiles such as the Navarro-Frenk-White \citep[NFW,][]{Navarro1996} profile, assuming a specific shape of the mass profile \citep[but see][]{johnston2007,Umetsu2011,Umetsu2025}.
An alternative is the non-parametric method from \citet{Mistele2024b} which does not presume a specific profile, allowing for less biased and more model-independent measurements.

In the following we will use this method to infer non-parametric mass profiles for galaxy clusters from the Cluster Lensing and Supernova Survey with Hubble \citep[CLASH,][]{Postman2012} and study their properties, including the relation between the dynamical and baryonic mass components.
In Sec.~\ref{sec:data}, we discuss the data we use with the methods described in Sec.~\ref{sec:method}.
We present our results in Sec.~\ref{sec:results} and after a brief discussion in Sec.~\ref{sec:discussion} we conclude in Sec.~\ref{sec:conclusion}.

\section{Data}
\label{sec:data}

We consider a subset of 20 galaxy clusters from the CLASH project for which weak-lensing observations, primarily from the Subaru Suprime-Cam, are available \citep{Umetsu2014}.
Of these, 16 were originally X-ray selected to be massive ($kT > 5\,\mathrm{keV}$, listed first in Table~\ref{tab:clusters}) and 4 were selected for their high lensing strength.

\subsection{Weak lensing data}

\begin{figure*}
\begin{center}
 \includegraphics[width=2\columnwidth]{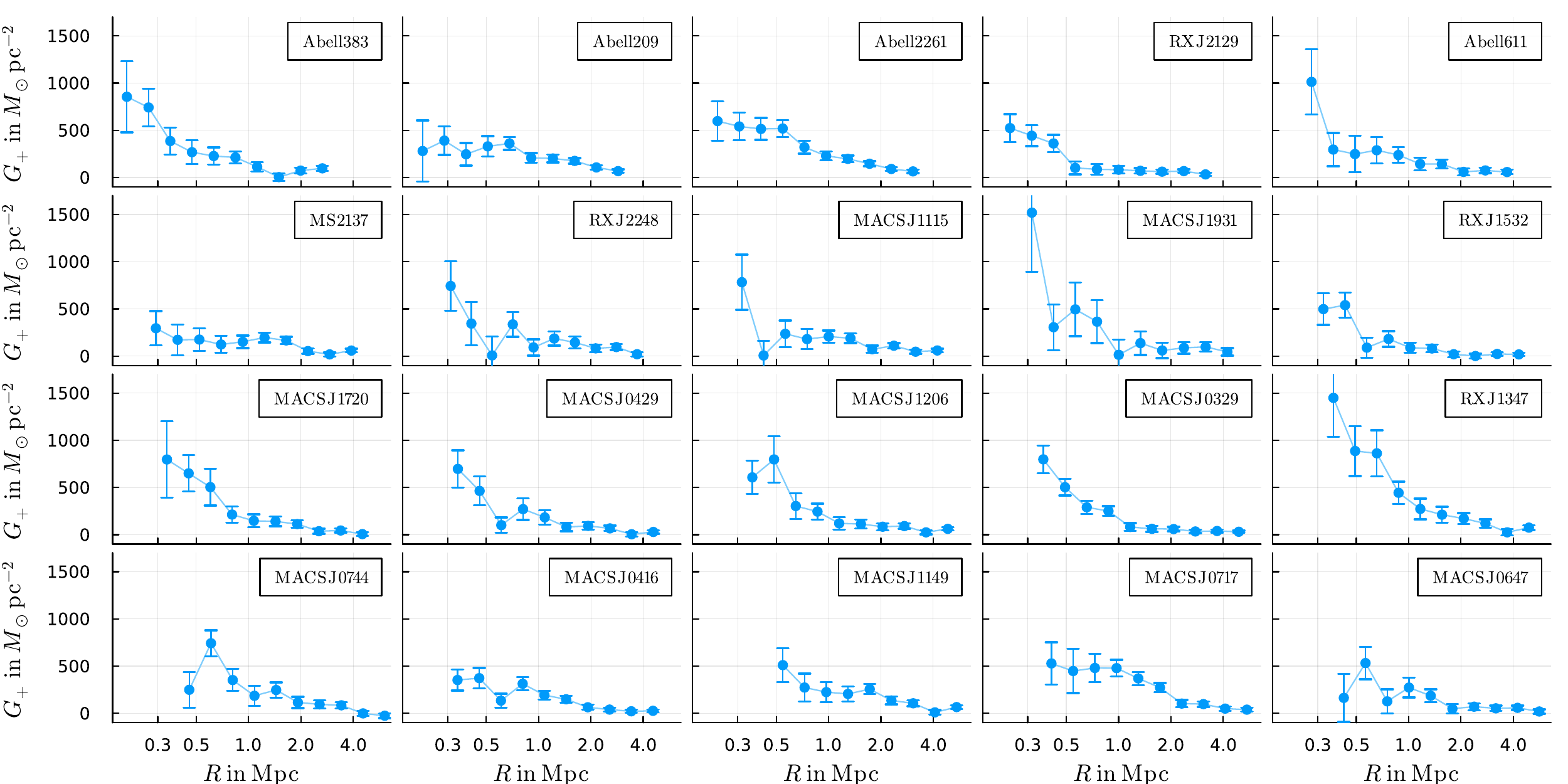}
\end{center}
\caption{The reduced shear in terms of $G_+ = \langle g_+ \rangle/\langle \Sigma_{\mathrm{crit}}^{-1} \rangle$ for the 20 CLASH clusters from \citet{Umetsu2014}.
 The error bars include contributions from the reduced shear $\langle g_+ \rangle$, the inverse critical surface density $\langle \Sigma_{\mathrm{crit}}^{-1} \rangle$, and from the LSS, see Sec.~\ref{sec:method:uncertainties}.}
\label{fig:G-grid}
\end{figure*}

We adopt the azimuthally-averaged reduced shear profiles $\langle g_+ \rangle$, cluster redshifts $z_l$, and the corresponding critical surface densities from \citet{Umetsu2014}.
The shear profiles are given as functions of angular distances, which we convert to projected radii $R$ assuming a flat $\Lambda$CDM cosmology with $H_0 = 70 \,\mathrm{km}\,\mathrm{s}^{-1}\,\mathrm{Mpc}^{-1}$ and $\Omega_m = 0.27$, following \citet{Umetsu2014}.
We adopt this cosmology throughout this work.

The critical surface density of a given lens-source pair is defined as,
\begin{equation}
\Sigma_{\mathrm{crit},ls}^{-1} = \frac{4 \pi G_{\mathrm{N}}}{c^2} \frac{D(z_l) D(z_l, z_s)}{D(z_s)} \equiv \frac{4 \pi G_N}{c^2} D(z_l) \cdot \beta \,,
\end{equation}
where $G_{\mathrm{N}}$ is the Newtonian gravitational constant and $D(z_l)$, $D(z_s)$, and $D(z_l, z_s)$ are the angular diameter distances to the lens, to the source, and between the source and the lens, respectively.
For our weak-lensing analysis below, we need $\langle \Sigma_{\mathrm{crit},ls}^{-1} \rangle$ and $\langle \Sigma_{\mathrm{crit},ls}^{-2} \rangle$ where $\langle \dots \rangle$ denotes averaging over source galaxies (see Sec.~\ref{sec:method:deprojection}).
\citet{Umetsu2014} provide estimates for $\langle \beta \rangle$ as well as $\langle \beta^2 \rangle / \langle \beta \rangle^2$ from which we infer the desired quantities using the cluster redshift $z_l$ and the assumed cosmology.

\begin{deluxetable}{lcc|cc}
\tablecaption{Properties of the CLASH clusters we use and quantities inferred from our weak-lensing analysis.}
\tablehead{
\colhead{Name} &
\colhead{$z_l$} &
\colhead{$R_{\mathrm{max}}^X$} &
\colhead{$\log_{10} M_{200c}^{\mathrm{2h\;sub}}$} &
\colhead{$\log_{10} M_{200c}$}
\\
\colhead{} &
\colhead{} &
\colhead{$\mathrm{Mpc}$} &
\colhead{$M_\odot$} &
\colhead{$M_\odot$}
}
\startdata
\input{./plots/table-cluster-list.tex}\enddata
\tablecomments{%
 $R_{\mathrm{max}}^X$ denotes the radial extent of the X-ray data used in \citet{Donahue2014}.
 $^\ast$The exception is RX J2129 where \citet{Famaey2024} redid the fit with a smaller radial range than the $R_{\mathrm{max}}^X$ listed in \citet{Donahue2014}.
 Missing $R_{\mathrm{max}}^X$ values indicate that a baryonic mass estimate from \citet{Famaey2024} is not available.
 The masses $M_{200c}$ are missing for Abell 383 due to the uptick in the non-parametric mass profile (see Fig.~\ref{fig:M-grid}).
}
\label{tab:clusters}
\end{deluxetable}

Our weak-lensing analysis makes use of the shear profiles in the form $G_+ = \langle g_+ \rangle/\langle \Sigma_{\mathrm{crit},ls}^{-1} \rangle$, which we show in Fig.~\ref{fig:G-grid}.
The shear $\langle g_+ \rangle$ varies as a function of projected radius while, following \citet{Umetsu2014}, $\langle \Sigma_{\mathrm{crit},ls}^{-1} \rangle$ is assumed to be constant within a given cluster.
The uncertainties shown in Fig.~\ref{fig:G-grid} are discussed in Sec.~\ref{sec:method:uncertainties}.

\subsection{Baryonic mass estimate}
\label{sec:data:Mb}

We adopt the baryonic mass estimates from \citet{Famaey2024}.
These are available for 16 out of 20 clusters in our sample (see Table~\ref{tab:clusters}) to which we restrict ourselves whenever baryonic masses are required.

\citet{Famaey2024} estimate the baryonic mass profile $M_b(r)$ as a sum of the dominant contribution from the intracluster medium, $M_{\mathrm{gas}}(r)$, and further contributions from the brightest cluster galaxy (BCG) including companions within $50\,\mathrm{kpc}$, $M_{\ast,\mathrm{BCG}^+}$, \citep{Burke2015} as well as contributions from other galaxies, $M_{\mathrm{gal}}(r)$.

Since we are only interested in radii larger than a few hundred $\mathrm{kpc}$, we treat $M_{\ast,\mathrm{BCG}^+}$ as a point mass.
The galaxy contribution is given as a fraction $f_{\mathrm{gal}}$ of $M_{\mathrm{gas}}$, i.e. $M_{\mathrm{gal}}(r) = f_{\mathrm{gal}}(r) \cdot M_{\mathrm{gas}}(r)$.
Following \citet{Famaey2024}, this fraction $f_{\mathrm{gal}}$ is assumed to be the same across all galaxy clusters when normalized to the radius $r_{200c}$\footnote{
 In the following, $M_{200c}$ refers to the mass within the radius $r_{200c}$ where the galaxy cluster's average mass density drops to 200 times the critical density at the cluster's redshift.
 } (that we measure from weak gravitational lensing, see below).%
 \footnote{
Specifically, \citet{Famaey2024} measure $f_{\mathrm{gal}}$ for MACS J1206 and adopt $f_{\mathrm{gal}}(r) = f_{\mathrm{gal},\mathrm{J}1206}\left(r \cdot r_{200c,\mathrm{J}1206}/r_{200c}\right)$ for the other clusters.
Beyond the last measured data point, $f_{\mathrm{gal},\mathrm{J}1206}$ is assumed to remain constant at the last measured value.
}
Values of $f_{\mathrm{gal}}$ typically reach $O(1)$ at $r \lesssim 0.1\,\mathrm{Mpc}$ and drop to about $8\%$ or $12\%$ beyond $r_{200c}$, depending on how we extrapolate the gas density profiles (see below).
This simple procedure doesn't allow precise quantitative statements about $M_b$, but suffices for our purposes.

The dominant contribution $M_{\mathrm{gas}} (r)$ is obtained by fitting double-beta profiles to X-ray observations,
\begin{multline}
 \rho_{\mathrm{gas}}(r)
  = n_0
    \left(\frac{r}{r_0}\right)^{-\alpha}
    \left(1 + \left(\frac{r}{r_{e,0}}\right)^2 \right)^{-\frac{3 \beta_0}{2}}
    \\
   + n_1 \left(1 + \left(\frac{r}{r_{e,1}}\right)^2 \right)^{-\frac{3 \beta_1}{2}}
   \,,
\end{multline}
with free parameters $n_0$, $r_0$, $\alpha$, $r_{e,0}$, $\beta_0$, $n_1$, $r_{e,1}$, and $\beta_1$.
We adopt the fit results from \citet{Famaey2024}.
Most of these are originally from \citet{Laudato2022} based on data from \citet{Donahue2014}, but \citet{Famaey2024} redid them for a few clusters for which the implied gas masses were unreasonably small.

The underlying Chandra X-ray observations to which these beta profiles were fit become noisy at large radii.
Therefore, \citet{Donahue2014} considered only data out to a maximum radius $R_{\mathrm{max}}^X$, which is determined by requiring at least $1500$ counts of photon signal in each radial bin. 
In the following, we assume that the beta profile fits are reliable up to $R_{\mathrm{max}}^X$\footnote{
 In the fits they redid, \citet{Famaey2024} did not impose the $1500$ photon count requirement and so used X-ray observations out to larger radii for some clusters.
 However, since this additional data is quite noisy, it still makes sense to adopt $R_{\mathrm{max}}^X$ from \citet{Donahue2014} as an estimate of how far out to trust the beta profile fits.
}.
At larger radii, beyond $R_{\mathrm{max}}^X$, we consider two options:
1) We assume that the best fit parameters remain valid even at large radii where they were not well constrained by observations and 2) we assume that the beta profiles are matched to a $1/r^4$ tail at $r = R_{\mathrm{max}}^X$.

The $1/r^4$ tail has two motivations.
If we require a finite total gas mass, we need a gas density that asymptotically decays faster than $1/r^3$ so $1/r^4$ is perhaps a natural choice.
Also, $1/r^4$ is the asymptotic behavior of isothermal spheres in MOND \citep{Milgrom1984}.

\begin{figure}
\includegraphics[width=\columnwidth]{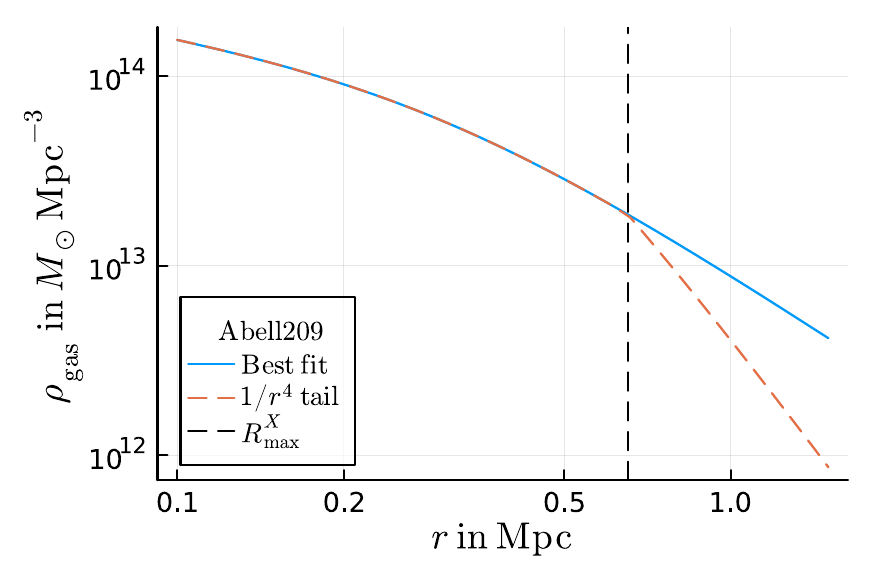}
\caption{%
 The gas density of Abell 209 implied by the double beta fit from \citet{Famaey2024}.
 The vertical dashed line indicates $R_{\mathrm{max}}^X$, i.e. how far out this fit is reliable. 
 We show two different ways of extrapolating beyond $R_{\mathrm{max}}^X$: Assuming the best-fit parameters to be valid even beyond $R_{\mathrm{max}}^X$ (solid blue line) and assuming a $1/r^4$ tail (dashed red line).
}
\label{fig:rhogas-a209}
\end{figure}

The $1/r^4$ matching procedure is illustrated in Fig.~\ref{fig:rhogas-a209}.
Beyond $R_{\mathrm{max}}^X$, we set $\beta_0 = (4 - \alpha)/3$ and $\beta_1 = 4/3$ to ensure an asymptotic $1/r^4$ decay and we adjust $n_0$ and $n_1$ such that the $n_0$ and $n_1$ components of $\rho_{\mathrm{gas}}$ are both continuous.
In practice, $R_{\mathrm{max}}^X$ is quite small for some clusters (see Table~\ref{tab:clusters}).
Thus, even if $1/r^4$ is the correct asymptotic behavior, one may worry that $R_{\mathrm{max}}^X$ is not yet in that asymptotic regime, so that matching to a $1/r^4$ tail at $R_{\mathrm{max}}^X$ may underestimate the true gas mass.
In that case one would expect our $1/r^4$ extrapolation to artificially induce a positive correlation between gas mass and $R_{\mathrm{max}}^X$.

\begin{figure}
\includegraphics[width=\columnwidth]{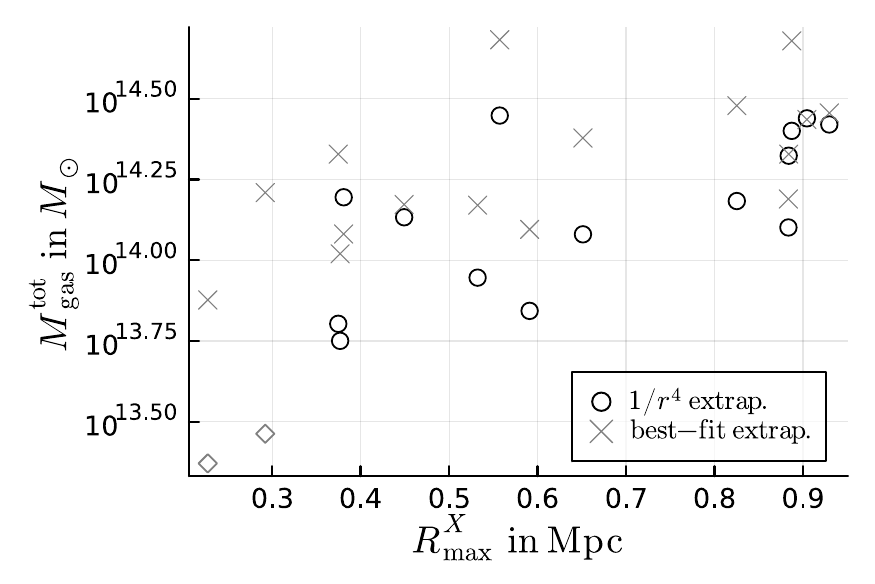}
\caption{%
 The radius $R_{\mathrm{max}}^X$ out to which the double beta profile fits of the gas densities are well constrained by observations versus the total gas mass. Circles correspond to assuming a $1/r^4$ density tail after $R_{\mathrm{max}}^X$.
 The two clear outliers, MS 2137 and MACS J0429, are shown as gray diamonds.
 Crosses correspond to assuming that the best-fit beta profiles remain valid even beyond $R_{\mathrm{max}}^X$.
}
\label{fig:rmax-vs-mgastot}
\end{figure}

To assess this effect, the white circles in Fig.~\ref{fig:rmax-vs-mgastot} show the total gas mass implied by our $1/r^4$ matching procedure versus the radial range of the X-ray observations $R_{\mathrm{max}}^X$.
For comparison, the gray crosses in Fig.~\ref{fig:rmax-vs-mgastot} correspond to assuming that the best-fit beta profiles remain valid even beyond $R_{\mathrm{max}}^X$ with the gas mass evaluated at $r = r_{200c}$ (see Sec.~\ref{sec:results:btfr}).
We see that assuming a $1/r^4$ tail induces only a weak additional correlation between the total gas mass and $R_{\mathrm{max}}^X$, though there are two clear outliers MS 2137 and MACS J0429, which have the smallest gas masses in our sample, and two borderline cases  MACS J0329 and MACS J1115, with the 3rd and 4th lowest gas masses.
Quantitatively, with our $1/r^4$ extrapolation, Pearson's $r$ is $0.82$ for the full sample and drops to $0.43$ with the two outliers and the two borderline cases removed.
With the best-fit extrapolation, Pearson's $r$ is $0.66$.
This suggests that, if $1/r^4$ is indeed the correct asymptotic behavior, it is overall not unreasonable to match to a $1/r^4$ tail at $R_{\mathrm{max}}^X$.
The gas masses for the two outliers are, however, likely to be significantly underestimated and we consider them separately below.
The gas masses of clusters with $R_{\mathrm{max}}^X$ similar to the two borderline cases may also be somewhat underestimated but we do not treat them separately.

Recently, the XMM Cluster Outskirts Project \citep[X-COP,][]{Eckert2017,Ghirardini2019} has constrained the gas densities of 12 galaxy clusters out to about $r_{200c}$ \citep[for earlier measurements at large radii using the Suzaku satellite, see][]{Walker2012,Walker2012b,Urban2014}.
They find density profiles that steepen with radius.
In fact, their density profiles become steeper than the asymptotic behavior of many of the best-fit beta profiles from \citet{Famaey2024}, though not as steep as $1/r^4$.
This suggests that the true asymptotic behavior of the gas profiles is in between the two extrapolations we use.
In any case, the point of adopting multiple different extrapolations is to illustrate a range of possibilities, not to give a precise quantitative result.

\section{Method}
\label{sec:method}

\subsection{Weak-lensing mass profile measurements}
\label{sec:method:deprojection}

Weak-lensing observations of a lens $l$, in our case a galaxy cluster, are based on the shapes of many background source galaxies $s$.
From these shapes and the position angle between the sources and the lens, one can infer the azimuthally-averaged tangential reduced shear $\langle g_t \rangle$ \citep{Bartelmann2001}, which encodes the cluster's projected mass distribution.
Indeed, assuming spherical symmetry, we have to a good approximation \citep{Umetsu2020,Mistele2024b}
\begin{equation}
 \label{eq:shear_ESD_relation}
 G_+ (R) = \frac{\Delta \Sigma (R)}{1 - f_c (R) \Sigma(R) } \,,
\end{equation}
where $R$ is the projected radius, $G_+$ is the azimuthally-averaged reduced tangential shear divided by the azimuthally-averaged inverse critical surface density,
\begin{equation}
 \label{eq:G+def}
 G_+ \equiv \frac{\langle g_+ \rangle}{\langle \Sigma^{-1}_{\mathrm{crit},ls} \rangle} \,,
\end{equation}
and $f_c$ refers to the following ratio of azimuthal averages of powers of the critical surface density,
\begin{equation}
 f_c = \frac{\langle \Sigma_{\mathrm{crit},ls}^{-2} \rangle}{\langle \Sigma_{\mathrm{crit},ls}^{-1} \rangle} \,.
\end{equation} 
Further, $\Sigma$ and $\Delta \Sigma$ are, respectively, the surface mass density and the excess surface mass density of the cluster. 
The excess surface density is defined as
\begin{equation}
 \label{eq:ESDdef}
 \Delta \Sigma (R) = \frac{M_{2D} (R)}{\pi R^2} - \Sigma (R)\,,
\end{equation}
where $M_{2D}(R)$ is the mass enclosed by a cylinder with radius $R$ that is oriented along the line of sight.

We will be interested in the deprojected, 3D mass profile $M(r)$.
To convert observations of the shear $G_+$ and the critical surface density $f_c$ into the 3D mass $M$, we assume spherical symmetry and use the deprojection technique of \citet{Mistele2024b} \citep[see also][]{Mistele2023d,Mistele2024}.
This technique is based on Eq.~\eqref{eq:shear_ESD_relation} and consists of two steps.
First, we convert $G_+$ and $f_c$ into an excess surface density $\Delta \Sigma$,
\begin{multline}
 \label{eq:ESD_from_shear}
 \Delta \Sigma (R) =
 \frac{G_+(R)}{1 - f_c G_+(R)}
 \\
 \times
 \exp\left(
  	    - \int_R^\infty dR' \frac{2}{R'} \frac{f_c G_+(R')}{1 - f_c G_+(R')}
       \right) \,.
\end{multline}
This step is only important at relatively small radii where the $f_c \Sigma$ term in the denominator of Eq.~\eqref{eq:shear_ESD_relation} is not negligible.
This roughly corresponds to $\Sigma/\Sigma_{\mathrm{crit}}$, also known as the convergence $\kappa$, not being negligible.
In contrast, at large radii where $\Sigma/\Sigma_{\mathrm{crit}}$ is small, Eq.~\eqref{eq:ESD_from_shear} reduces to $G_+ = \Delta \Sigma$.
The second step is to convert $\Delta \Sigma (R)$ from Eq.~\eqref{eq:ESD_from_shear} into the 3D mass profile $M(r)$,
\begin{equation}
 \label{eq:M_from_ESD}
 M(r) = 4 r^2 \int_0^{\pi/2} d \theta \, \Delta \Sigma \left( \frac{r}{\sin \theta}\right) \,.
\end{equation}

Equation \eqref{eq:ESD_from_shear} is mathematically valid as long as $G_+ f_c < 1$.
This roughly corresponds to the condition that we are in the weak-lensing regime \citep{Mistele2024b}.
Equation \eqref{eq:ESD_from_shear} further assumes that the critical surface density, i.e. $f_c$, is constant as a function of projected radius $R$.
\citet{Mistele2024b} also provide formulas for the case with a radially varying $f_c$.
A constant $f_c$ is, however, often a reasonable assumption in practice and, following \citet{Umetsu2014}, we adopt that assumption here.

For the integrals in the above deprojection formulas, we need to know the function $G_+(R)$ at all radii out to infinity, whereas, in practice, we measure $G_+$ only in a discrete set of radial bins out to some outermost bin with bin center $R_{\mathrm{max}}$.
Thus, we must interpolate between the discrete radial bins and extrapolate beyond $R_{\mathrm{max}}$.
In practice, the effect of the interpolation on the inferred $M(r)$ is quite minor and the extrapolation is unimportant except when $r$ is close to the last measured data point at $R_{\mathrm{max}}$.
As a result, the choices we make about how to extrapolate and interpolate are unimportant over much of the radial range we consider \citep{Mistele2024b,Mistele2023d,Mistele2024}.

To be specific, we extrapolate assuming that $G_+$ follows a $1/R$ power law beyond $R_{\mathrm{max}}$ and we linearly interpolate the discrete $G_+$ measurements in logarithmic space.
We take the uncertainty in these choices into account as systematic uncertainties, see Sec.~\ref{sec:method:uncertainties}.
In Appendix~\ref{sec:appendix:nfwextrapolation}, we show that extrapolating $G_+$ assuming an NFW profile instead of a $1/R$ power law does not significantly change our results, further confirming that our systematic uncertainty estimate is reasonable.

The above procedure is straightforward to implement numerically and runs fast, taking only a few milliseconds per galaxy cluster.
We use the code provided by \citet{Mistele2024b}\footnote{\url{https://github.com/tmistele/SphericalClusterMass.jl}} to implement Eq.~\eqref{eq:ESD_from_shear} and Eq.~\eqref{eq:M_from_ESD} as well as the propagation of uncertainties and covariances discussed in Sec.~\ref{sec:method:uncertainties} below.

\subsection{Two-halo term}
\label{sec:method:twohalo}

The deprojection technique discussed in Sec.~\ref{sec:method:deprojection} assumes that all of the lensing signal is due to the cluster itself.
This is a good approximation at small and moderate radii.
Beyond a few $\mathrm{Mpc}$, however, the signal from the cluster's local environment, the so-called two-halo term, can become important \citep{Umetsu2020,Oguri2011,Oguri2011b}.

Within $\Lambda$CDM, we can estimate this contribution and subtract it.
The resulting subtracted mass profiles should be a better estimate of the true mass profiles, but are more model-dependent.
Below we present results with and without this two-halo subtraction, but the difference turns out not to be important for our purposes.
We explain the details of our two-halo subtraction procedure in Appendix~\ref{sec:appendix:twohalo}.

\subsection{Uncertainties and covariances}
\label{sec:method:uncertainties}

We consider two sources of systematic uncertainties, corresponding to our choices of how to extrapolate $G_+$ beyond the last measured data point and of how to interpolate between the discrete radial bins.
For the statistical uncertainties, we take into account uncertainties in the shear measurements $\langle g_+ \rangle$, as well as uncertainties and covariances from the inverse critical surface densities  $\langle \Sigma_{\mathrm{crit}}^{-1} \rangle$, and covariances induced by the large-scale structure \citep[LSS,][]{Hoekstra2003}.
We use linear error propagation to propagate the statistical uncertainties and covariances into the reconstructed mass profiles.
As a result, any shortcomings in our uncertainty and covariance estimates affect only the error bars of the reconstructed mass profiles; they do not enter the central values.
The details are discussed in Appendix~\ref{sec:appendix:uncertainties}.

\subsection{Non-parametric density reconstruction}
\label{sec:method:densitydeprojection}

One way to obtain a 3D density profile $\rho(r)$ in a non-parametric way is to first reconstruct the 3D mass profile  $M(r)$ following Sec.~\ref{sec:method:deprojection} and then take a numerical derivative, $\rho(r) = M'(r)/4 \pi r^2$.
However, numerical derivatives can be tricky.
Thus, we here introduce a density reconstruction method that goes directly from shear to density without any numerical derivatives.

Specifically, as we show in Appendix~\ref{sec:appendix:density}, assuming spherical symmetry we have\footnote{
 The factor of $1/\pi$ may be reminiscent of the Abel transform, $\rho(r) = - \pi^{-1} \int_r^\infty dR \, \Sigma'(R)/\sqrt{R^2 - r^2} $.
 Indeed, the Abel transform can be written in an alternative way that looks quite similar to Eq.~\eqref{eq:densityreconstruction} and contains no derivatives of $\Sigma$: $\rho(r) = - (\pi r)^{-1} \int_0^{\pi/2} d\theta \, \frac{1}{\cos^2 \theta} \left(\Sigma\left(\frac{r}{\sin \theta}\right) - \Sigma(r)\right)$ .
}
\begin{equation}
 \label{eq:densityreconstruction}
 \rho(r) = \frac{I_1(r) + I_2(r)}{\pi r} \,,
\end{equation}
with
\begin{align}
 \label{eq:densityreconstruction:I1}
 I_1(r) &= \int_0^{\pi/2} d\theta \,
  \frac{\Delta \Sigma \left(\frac{r}{\sin \theta}\right) - \Delta \Sigma(r)}{\cos^2 \theta} \,, \\
 I_2(r) &= 2 \int_0^{\pi/2} d\theta \, \Delta \Sigma\left(\frac{r}{\sin \theta}\right) \,.
\end{align}
This replaces the second step Eq.~\eqref{eq:M_from_ESD} in the reconstruction method from Sec.~\ref{sec:method:deprojection}, i.e. the step that converts $\Delta \Sigma$ to $M$.
The first step, Eq.~\eqref{eq:ESD_from_shear}, which converts the shear $G_+$ to $\Delta \Sigma$, remains the same.
The integrand of Eq.~\eqref{eq:densityreconstruction:I1} is finite everywhere despite the $1/\cos^2 \theta$ factor.
This can be seen by expanding around $\theta = \pi/2$.

While this non-parametric density reconstruction does not involve numerical derivatives, it still requires relatively low-noise data for good results due to the finite difference $\Delta \Sigma(r/\sin \theta) - \Delta \Sigma(r)$ in the numerator in Eq.~\eqref{eq:densityreconstruction:I1}.
Indeed, the $1/\cos^2 \theta$ up-weights precisely the region $\theta \approx \pi/2$ of the integral where $\Delta \Sigma(r/\sin \theta)$ and $\Delta \Sigma(r)$ are close to each other so that the difference becomes small.
Despite this, initial tests with mock data show that, at least in some situations, the non-parametric density reconstruction method Eq.~\eqref{eq:densityreconstruction} is better behaved than taking a numerical derivative of the non-parametric mass profile from Sec.~\ref{sec:method:deprojection}.

It should be possible to adjust the two-halo subtraction procedure from Sec.~\ref{sec:method:twohalo} to work with this non-parametric density reconstruction method.
This is left for dedicated follow-up work.

\subsection{On the assumption of spherical symmetry}
\label{sec:method:triaxial}

Our non-parametric deprojection method assumes spherical symmetry.
As we will now discuss, this assumption can be relaxed in some important ways, making the method more widely applicable and more robust.

First, consider clusters that are approximately spherically symmetric at large radii but whose inner regions are more complex.
Examples may be late-stage mergers or clusters with a complex baryonic mass distribution.
As expected, our deprojection method does not work well in the inner regions of such clusters due to the lack of symmetry.
However, perhaps surprisingly, the method does infer the correct enclosed mass at larger radii, outside the non-symmetric core \citep{Mistele2024b}.
Thus, unlike methods based on fitting NFW or similar profiles, our method will not be thrown off by complexity or non-symmetry at small radii.

A related useful property of our deprojection method is that mass measurements at large radii (beyond a few times the miscentering offset) are not affected much by miscentering.
This is because the effect of miscentering is essentially to induce non-symmetry at small radii, while keeping approximate spherical symmetry at large radii.
In addition, there is a way to efficiently correct for residual miscentering effects.
Below, we do not consider any miscentering effects, so for brevity we refer the reader to \citet{Mistele2024b} for details.

\begin{figure}
\includegraphics[width=\columnwidth]{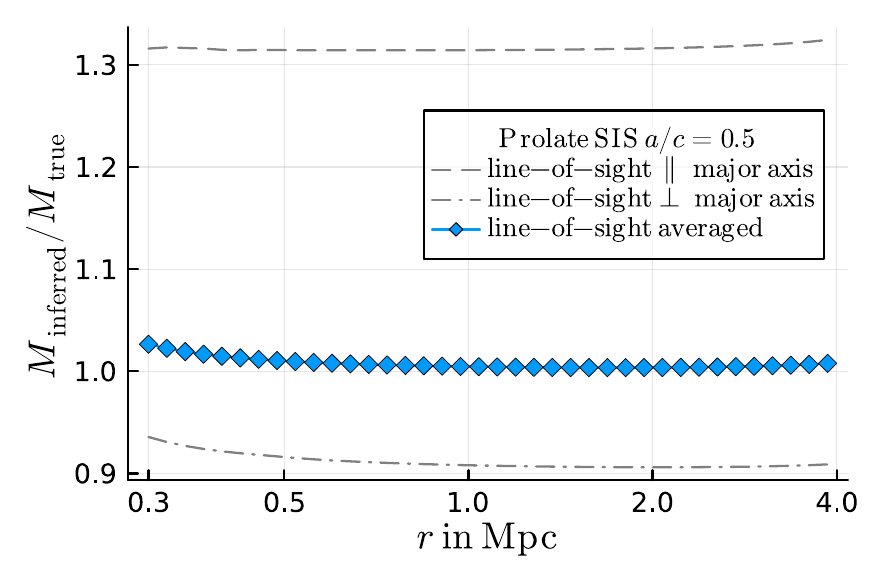}
\caption{%
  The mass profile of a prolate SIS, inferred from its reduced shear using our non-parametric deprojection method (Sec.~\ref{sec:method:deprojection}), relative to the true mass profile (see Eq.~\eqref{eq:Mtrue}) for different orientations of the line of sight.
   When averaged over all line of sight orientations, the inferred mass matches the true mass at large radii where $\Sigma/\Sigma_{\mathrm{crit}}$ is negligible.
   At small radii, the non-linearity due to $\Sigma/\Sigma_{\mathrm{crit}}$ (see Eq.~\eqref{eq:shear_ESD_relation}) induces a small deviation from unity.
   Our choice of $\Sigma_{\mathrm{crit}}$ maximizes this non-linear effect.
   The tiny but perceptible uptick at the largest radii shown is an edge effect (see the main text).
 }
\label{fig:prolate-sis-demo}
\end{figure}

Finally, as we will now argue, our method works well, on average, even for triaxial mass distributions such as dark matter halos in $\Lambda$CDM \cite[e.g.,][]{Bonamigo2015,Jing2002}.
The important caveat here is ``on average''.
Indeed, the mass inferred by our deprojection method for an individual triaxial halo can be off by a few $10\%$.
This is illustrated by the gray lines in Fig.~\ref{fig:prolate-sis-demo} for a prolate singular isothermal sphere (SIS).

Concretely, Fig.~\ref{fig:prolate-sis-demo} assumes a 3D density $\rho(x, y, z) = \rho_{\mathrm{SIS}}(\sqrt{(x'^2+y'^2)/a^2+z'^2/c^2})$ where the coordinates $x, y, z$ and $x', y', z'$ are related by a rotation that determines the orientation of the line-of-sight and where $\rho_{\mathrm{SIS}}(r) \propto 1/r^2$.
We use the formulas from \citet{Tessore2015} to calculate the reduced shear assuming, for simplicity, a single source plane with a constant $\Sigma_{\mathrm{crit}} = 1750\,M_\odot/\mathrm{pc}^2$.
We adopt $c = \sqrt{2}$, $a = 1/\sqrt{2}$, and choose the prefactor of $\rho_{\mathrm{SIS}}$ such that, with $c=a=1$, the mass within $1\,\mathrm{Mpc}$ is $10^{15}\,M_\odot$.
The inferred mass $M_{\mathrm{inferred}}$ is calculated by applying the deprojection formulas Eq.~\eqref{eq:ESD_from_shear} and Eq.~\eqref{eq:M_from_ESD} to the reduced shear. 
The true mass $M_{\mathrm{true}}$ is calculated from Eq.~\eqref{eq:Mtrue} (see below).
Our choice of $\Sigma_{\mathrm{crit}}$ is as small as possible, given that we are assuming \emph{weak} lensing (we enforce $G_+ f_c < 1$ in the radial range of Fig.~\ref{fig:prolate-sis-demo}, see Sec.~\ref{sec:method:deprojection}).
This maximizes the non-linear effect at small radii discussed below.

\begin{figure*}
\begin{center}
\includegraphics[width=2\columnwidth]{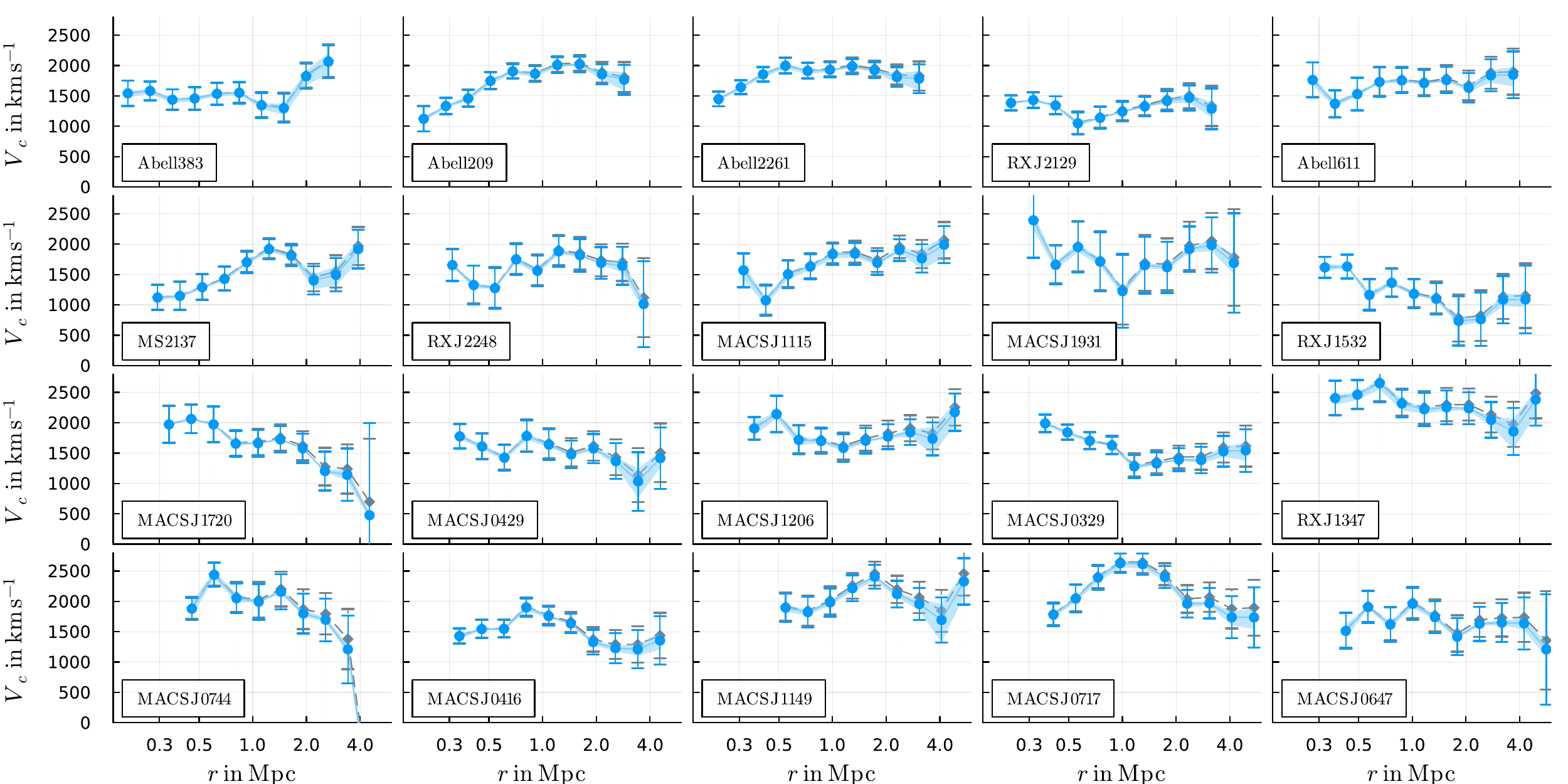}
\end{center}
\caption{%
 The circular velocities $V_c = \sqrt{G M(r)/r}$ inferred from the shear profile shown in Fig.~\ref{fig:G-grid} using our non-parametric deprojection method (Sec.~\ref{sec:method:deprojection}).
 Results after subtracting the two-halo term are in blue, unsubtracted results are in gray.
 The two-halo subtraction is based on $M_{200c}$ (see Sec.~\ref{sec:method:twohalo}) for all clusters except Abell 383, where we use $M_{500c}$ instead, because no value for $M_{200c}$ can be determined due to the uptick at large radii.
 The light blue band indicates the systematic uncertainties due to choices in how to interpolate and extrapolate the shear profiles (Sec.~\ref{sec:method:uncertainties}).
 Error bars indicate the statistical uncertainties.
}
\label{fig:vc-grid}
\end{figure*}

The inferred mass $M_{\mathrm{inferred}}$ depends on how the halo is oriented relative to the line of sight (gray lines in Fig.~\ref{fig:prolate-sis-demo}).
However, after averaging over all line of sight orientations, $M_{\mathrm{infered}}$ becomes very close to the true mass $M_{\mathrm{true}}$ (see the blue symbols in Fig.~\ref{fig:prolate-sis-demo}), where $M_{\mathrm{true}}$ is defined as the mass enclosed in a sphere with radius $r$,
\begin{equation}
 \label{eq:Mtrue}
 M_{\mathrm{true}} (r) \equiv \int_{|\vec{x}'| < r} d^3\vec{x}' \, \rho(\vec{x}') \,.
\end{equation}
In fact, at sufficiently large radii, the line of sight average of $M_{\mathrm{inferred}}$ is \emph{exactly} $M_{\mathrm{true}}$.
Sufficiently large radii here means radii where $\Sigma/\Sigma_{\mathrm{crit}} \ll 1$, which is usually a good approximation beyond a few hundred $\mathrm{kpc}$.
We prove this result in Appendix~\ref{sec:appendix:los-average} and show that it holds for \emph{any} mass distribution, not just triaxial ones (in the radial range where $\Sigma/\Sigma_{\mathrm{crit}} \ll 1$).
The proof only requires one mild and reasonable additional assumption on the redshift distribution of the source galaxy population.
At small radii, line-of-sight averaging does not exactly recover the true mass profile due to the non-linearity induced by $\Sigma/\Sigma_{\mathrm{crit}}$ (see Eq.~\eqref{eq:shear_ESD_relation}), but that effect seems to be quite mild in practice (Fig.~\ref{fig:prolate-sis-demo}).

Thus, on average, we expect our inferred mass profiles to be very close to the true mass profiles even for triaxial halos.
This is important for statistical analyses of large samples of galaxy clusters, for example for cluster cosmology.
This is particularly true for analyses based on quantities like $M_{200c}$, since these correspond to relatively large radii where $\Sigma/\Sigma_{\mathrm{crit}}$ is small.
This result may in principle change if the line-of-sight average is incomplete, for example due to intrinsic alignments or due to selection effects.
We expect such effects to be relatively minor, but leave a detailed study for future work.

Figure~\ref{fig:prolate-sis-demo} shows a tiny but perceptible deviation from unity at the largest radii.
This is an edge effect due to our choice of extrapolating $G_+$ beyond the last data point assuming a $1/R$ decay (see Sec.~\ref{sec:method:deprojection}) which is not exactly true here. %
Such effects are taken into account in our systematic error estimate (see Sec.~\ref{sec:method:uncertainties}).

\section{Results}
\label{sec:results}

\subsection{Circular velocities and total mass profiles}
\label{sec:results:vc}

\begin{figure*}
\begin{center}
\includegraphics[width=2\columnwidth]{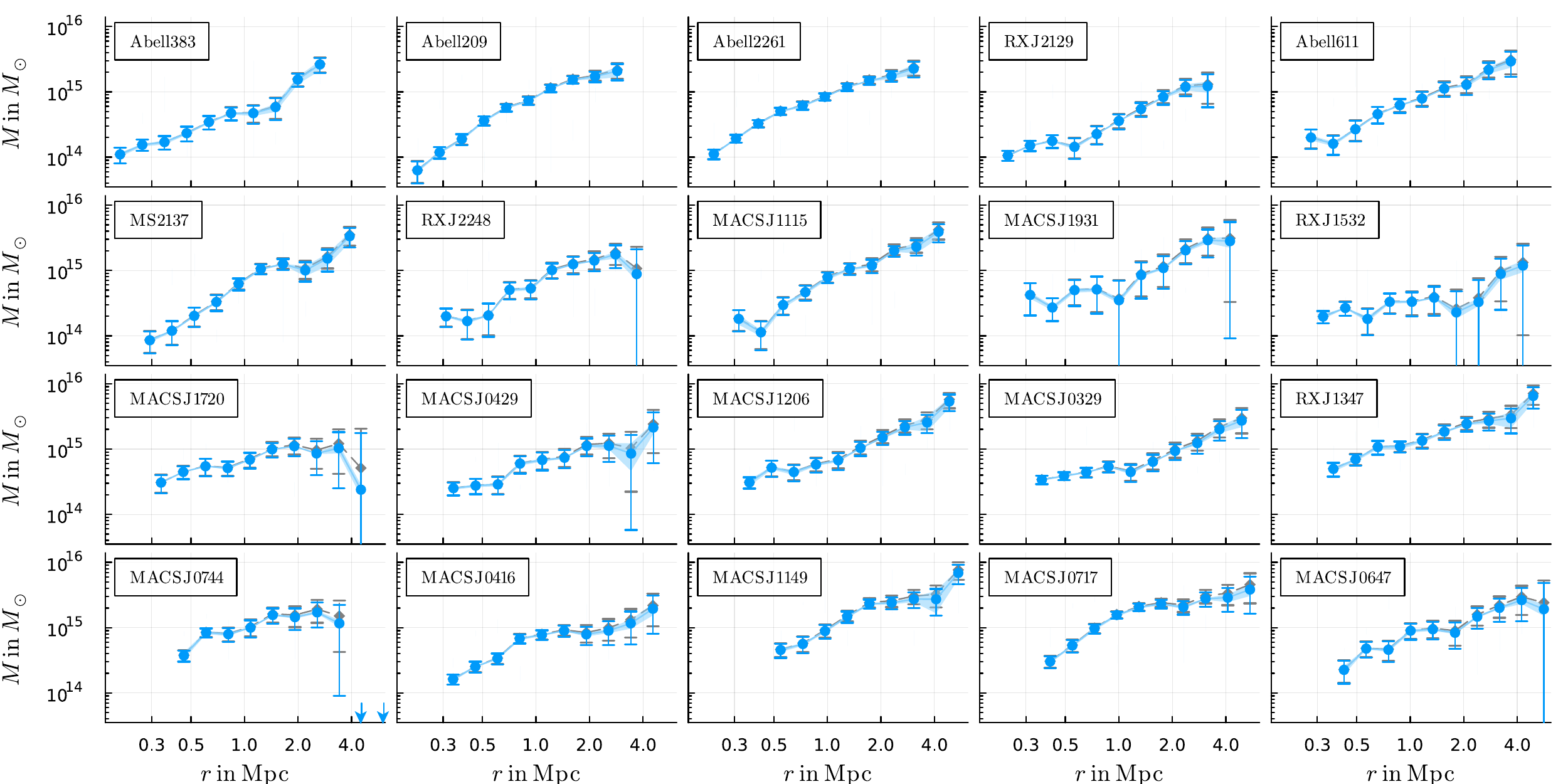}
\end{center}
\caption{%
 Same as Fig.~\ref{fig:vc-grid} but in terms of masses $M(r)$.
 Arrows at the horizontal axis indicate negative inferred masses.
}
\label{fig:M-grid}
\end{figure*}

Figure~\ref{fig:vc-grid} and Fig.~\ref{fig:M-grid} show the non-parametric mass profiles inferred using our non-parametric deprojection method in terms of the implied circular velocities $V_c (r) = \sqrt{G M(r) /r}$ and in terms of $M(r)$, respectively.
We show results with the $\Lambda$CDM two-halo contribution subtracted (blue symbols, see Sec.~\ref{sec:method:twohalo}) and without this subtraction (gray symbols).
The light blue band indicates systematic uncertainties due to having to interpolate and extrapolate the observed shear profiles (Sec.~\ref{sec:method:uncertainties}).
They become important only close to the last data point.
Neighboring data points in Fig.~\ref{fig:vc-grid} and Fig.~\ref{fig:M-grid} are correlated (see Sec.~\ref{sec:method:uncertainties}, Appendix~\ref{sec:appendix:uncertainties}, and Fig.~\ref{fig:a209-correlation}).

The circular velocities are remarkably flat, with no clear indication of a decline at large radii.
This is reminiscent of galaxy rotation curves and perhaps indicates a universal pattern.
The approximate flatness was previously noted in \citet{Donahue2014}.
In contrast to our analysis, \citet{Donahue2014} assumed a parametric NFW profile and did not subtract the two-halo contribution.
Asymptotically, the circular velocities implied by NFW halos decay like $\sqrt{\ln(r)/r}$ and so we should not expect to see asymptotically flat circular velocities in $\Lambda$CDM.
However, due to the relatively small concentrations of the dark matter halos of massive galaxy clusters and the limited radial range probed, our results are not very sensitive to this expected asymptotic decline.

\begin{figure}
\includegraphics[width=\columnwidth]{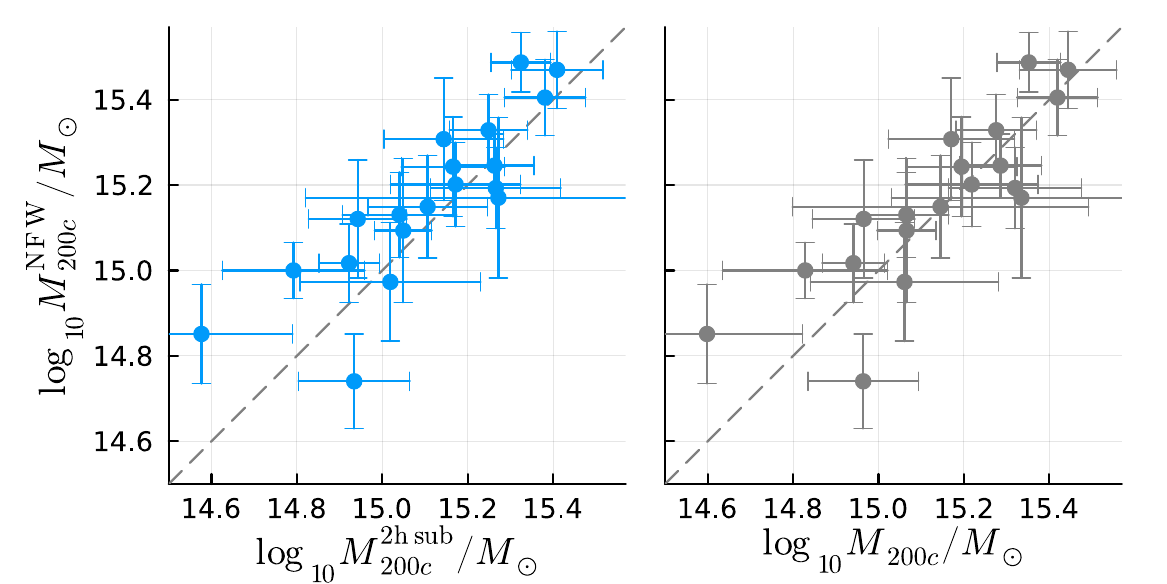}
\caption{%
 Our non-parametric $M_{200c}$ measurements compared to the $M_{200c}$ implied by the NFW fits from \citet{Umetsu2014} with (left) and without (right) the two-halo term subtracted (Sec.~\ref{sec:method:twohalo}).
 The NFW fits from \citet{Umetsu2014} do not take into account the two-halo term.
 Abell 383 is excluded because no value of $M_{200c}$ could be determined (see Fig.~\ref{fig:M-grid}).
 Error bars indicate statistical uncertainties.
}
\label{fig:M200scatter}
\end{figure}

In terms of the total mass $M_{200c}$, our non-parametric method gives results that are compatible with the NFW fits from \citet{Umetsu2014}.
There is a small and expected shift towards smaller masses (Fig.~\ref{fig:M200scatter}, left, see also Table~\ref{tab:clusters}) because \citet{Umetsu2014} did not take into account the two-halo term.
Indeed, if we turn off our two-halo subtraction procedure, there is no longer a significant shift (Fig.~\ref{fig:M200scatter}, right).
Figure~\ref{fig:M200scatter} also shows that our uncertainties on $M_{200c}$ are only moderately larger than those from \citet{Umetsu2014}, despite making significantly fewer assumptions.
On average, our statistical uncertainties are larger by $36\%$ in the case without two-halo subtraction.
This average and Fig.~\ref{fig:M200scatter} do not include Abell 383 because we could not determine a value of $M_{200c}$ due to the uptick at large radii (see Fig.~\ref{fig:M-grid}).
 
\begin{figure}
\includegraphics[width=.9\columnwidth]{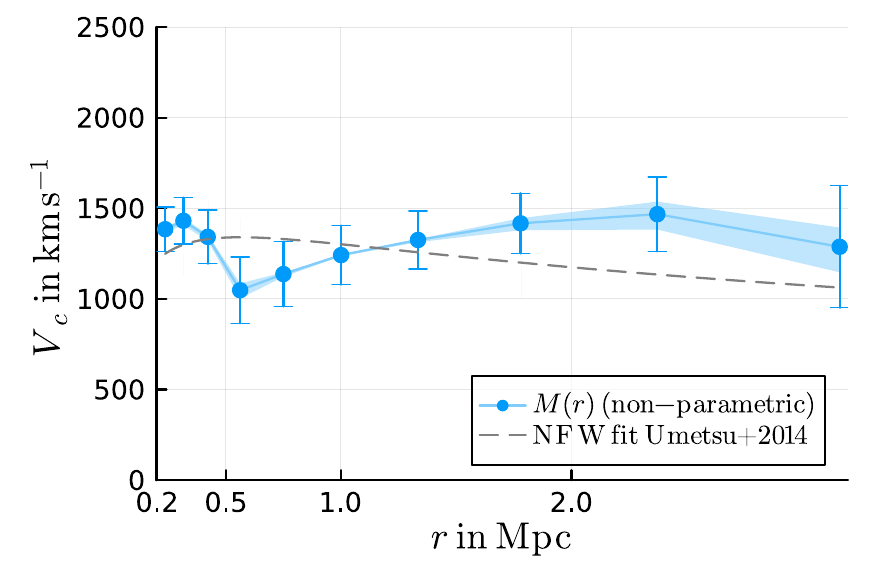}
\caption{%
 The circular velocity of RX J2129 as in Fig.~\ref{fig:vc-grid}, with the two-halo term subtracted, (blue) and the circular velocity implied by the NFW fit from \citet{Umetsu2014} (gray).
 A similar comparison for the other clusters can be found in Fig.~\ref{fig:vc-nfw-grid} in Appendix~\ref{sec:appendix:nfwextrapolation}.
}
\label{fig:rxj2129-nfw}
\end{figure}

Nevertheless, an NFW profile is not a great fit for all clusters.
For example, the NFW fit from \citet{Umetsu2014} for RX J2129 seems to indicate a clearly declining circular velocity beyond $\sim 0.5\,\mathrm{Mpc}$ \citep[see also Fig.~\ref{fig:rxj2129-nfw}]{Donahue2014}.
In contrast, our non-parametric mass profile implies a monotonic rise in the range $(0.5-2)\,\mathrm{Mpc}$.
Inspection of Fig.~\ref{fig:rxj2129-nfw} and the corresponding shear profile in Fig.~\ref{fig:G-grid} suggests that the NFW fit is thrown off by the data points at small radii, where the shear profile (and our non-parametric mass profile) show a clear change of behavior.
The reason for this qualitative change in behavior may simply be statistical fluctuations or it may be real complexity in the lens' mass distribution that is not captured by the simple NFW model.
RX J2129 is classified as relaxed \citep{Donahue2016}, but this does not necessarily preclude any significant non-symmetric structures in its center.
For example, the temperature profile is not well described by an isothermal profile \citep{Jimenez-Teja2024} and, using additional strong- and weak-lensing data from Hubble, \citet{Merten2015} noted some interesting morphology in the core of RX J2129.
The assumption of spherical symmetry may also be violated in other clusters which are known ongoing mergers with complex X-ray emission, for example MACS J0717 \citep{Ma2009} and MACS J0416 \citep{Mann2012}.

\begin{figure*}
\begin{center}
\includegraphics[width=2\columnwidth]{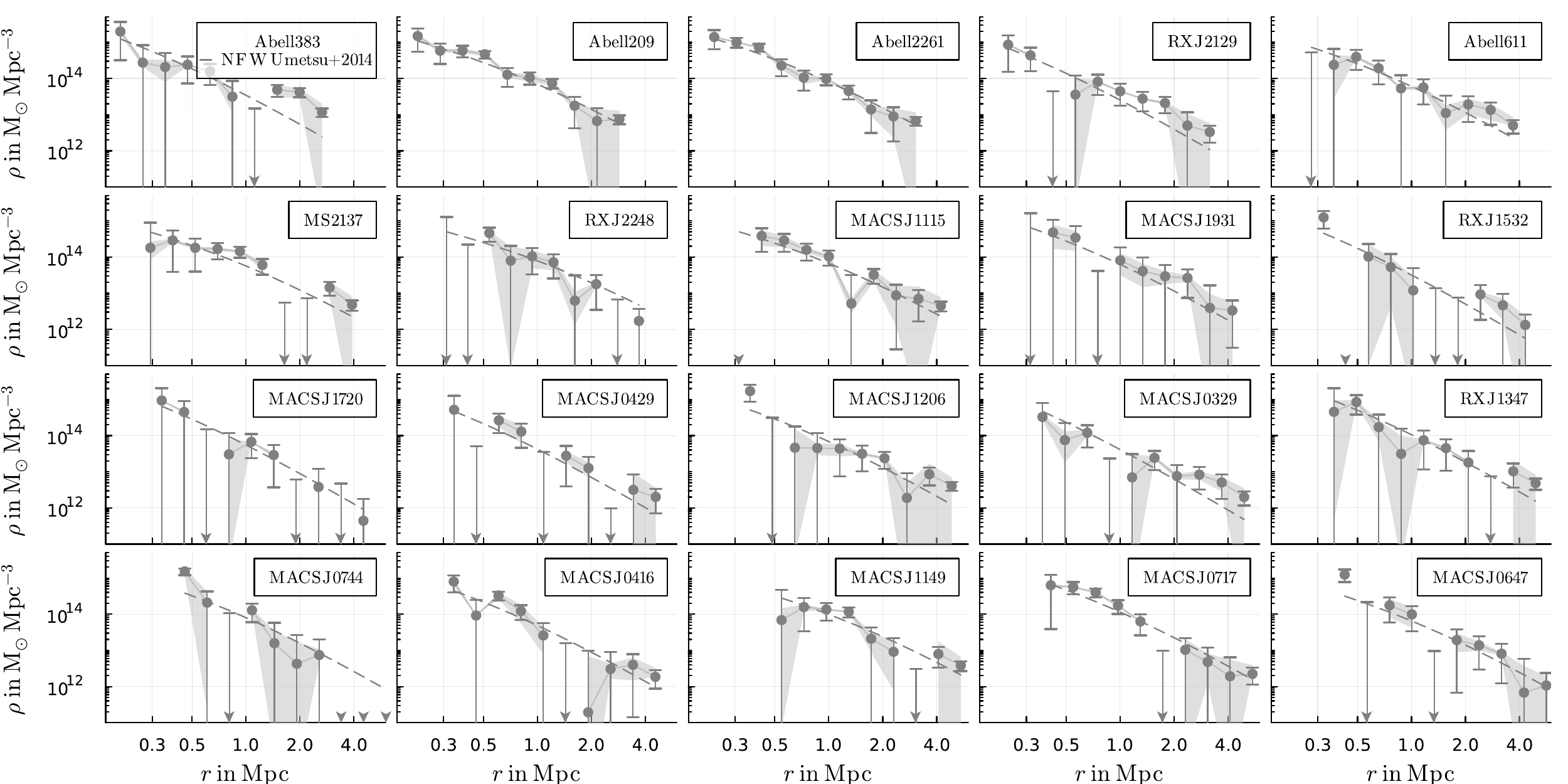}
\end{center}
\caption{%
 The 3D density profiles inferred using the non-parametric deprojection method from Sec.~\ref{sec:method:densitydeprojection}.
 For simplicity, no two-halo term is subtracted.
 Negative inferred densities are indicated by arrowheads at the horizontal axis.
 As in Fig.~\ref{fig:vc-grid}, light bands indicate systematic uncertainties from extrapolating and interpolating the shear profiles.
 Error bars indicate statistical uncertainties.
  Dashed gray lines show the densities implied by the NFW fits from \citet{Umetsu2014}.
}
\label{fig:rho-grid}
\end{figure*}

In any case, this highlights an important advantage of our non-parametric method:
Its mass estimates at large radii are not thrown off by complexity in the inner regions of a galaxy cluster.
This is discussed in detail in Sec.~\ref{sec:method:triaxial} and in \citet{Mistele2024b}.
To be clear, if the mass distribution in the inner regions of, e.g., RX J2129 is highly non-symmetric, then our non-parametric method cannot be trusted either at these small radii.
However, unlike an NFW fit, our inferred mass at larger radii will still be correct in such cases.

In contrast to NFW halos, asymptotically flat circular velocities are a prediction of alternative proposals such as MOND and, to varying degrees \citep{Mistele2023,Mistele2023c,Durakovic2023}, its relativistic completions \cite[e.g.][]{Skordis2020,Blanchet2024,Berezhiani2015}.
Qualitatively, our circular velocities are consistent with that prediction.
However, the precise predictions of these models depend strongly on the baryonic mass distribution.
Previous studies find that MOND-like theories underpredict the observed circular velocities of galaxy clusters, given their baryonic mass \citep[for some recent works see, e.g.,][]{Li2023,Li2024b,Famaey2024,Tian2020b,Kelleher2024,Ettori2019,Eckert2022}.
That said, the fact that the observed circular velocities of clusters seem to be approximately flat enables a potential solution to this discrepancy in terms of a missing baryonic mass component located at relatively small radii.
We discuss this more in Sec.~\ref{sec:results:missing}.

We note that our two-halo subtraction procedure is specific to $\Lambda$CDM and may not apply in other theories.
Unfortunately, due to their inherent non-linearity \citep[but see][]{Milgrom2025}, reliably estimating the two-halo contribution in MOND-inspired theories is challenging.
To the best of our knowledge, no such estimates are currently available.
One specific non-linear effect that would have to be taken into account is the so-called external field effect \citep{Bekenstein1984,Haghi2016,Chae2020,Chae2021b} which may play a role at large radii in galaxy clusters \citep{Kelleher2024}.
Since the effect of the $\Lambda$CDM two-halo term on our results  is quite modest, we might expect the same to be true for MOND-like theories.
Properly addressing this question will, however, require simulations of structure formation in relativistic models such as AeST.

\subsection{Density profiles}
\label{sec:results:density}

Figure~\ref{fig:rho-grid} shows the 3D density profiles inferred using the non-parametric method from Sec.~\ref{sec:method:densitydeprojection}.
This method is mathematically equivalent to taking the derivative of the non-parametric mass profiles $M(r)$ from Sec.~\ref{sec:method:deprojection} and dividing by $4 \pi r^2$.
The integral form Eq.~\eqref{eq:densityreconstruction} may, however, be preferable in practice because it avoids numerical derivatives.
Nevertheless, reconstructing good density profiles requires much smaller statistical uncertainties than reconstructing good mass profiles.
Indeed, for some clusters, such as Abell 209 or Abell 2261, the reconstructed densities look reasonable, but Fig.~\ref{fig:rho-grid} also shows large fluctuations and even negative inferred densities.

That said, while the negative densities in Fig.~\ref{fig:rho-grid} are likely just fluctuations, our deprojection method \emph{can} properly handle negative densities.
This is important for theoretical models where negative densities are real physical effects.
Examples are the ghost condensate in AeST \citep{Mistele2023} and the ``phantom dark matter'' in some modified gravity models \citep{Milgrom1986b}.

\begin{figure*}
\begin{center}
\includegraphics[width=2\columnwidth]{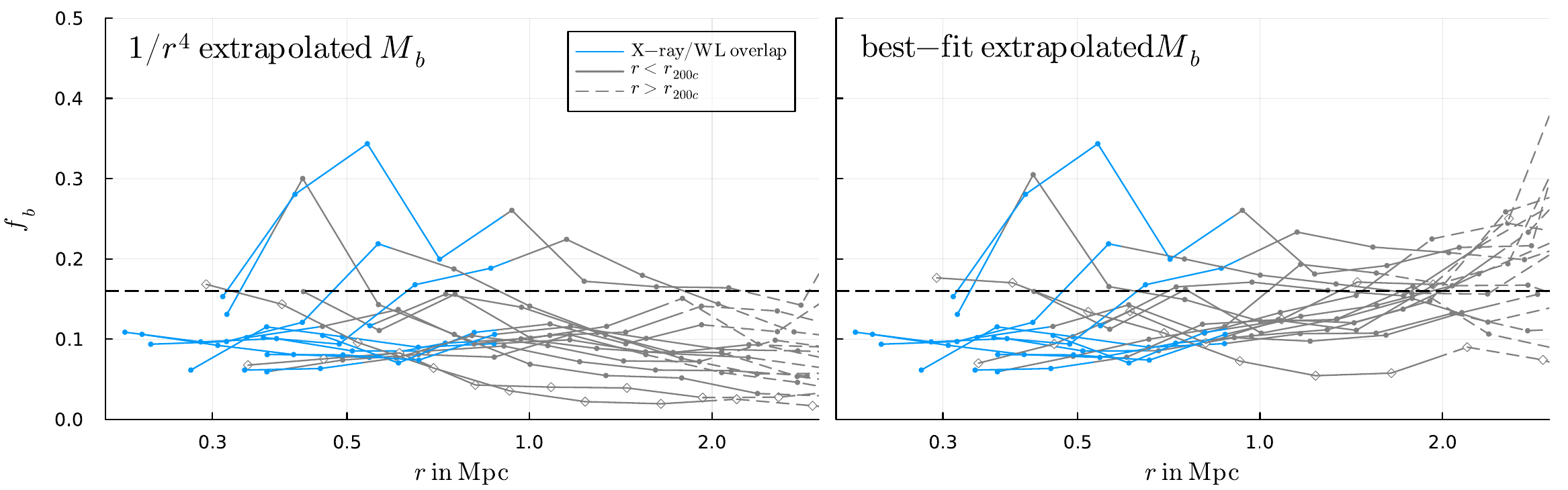}
\end{center}
\caption{%
  The baryon fraction $M_b (r) / M(r)$ implied by our non-parametric mass profiles as a function of radius.
  Blue lines indicate the radial range where weak-lensing and X-ray observations overlap, solid gray lines indicate radii beyond that but below $r_{200c}$, and dashed gray lines indicate radii beyond $r_{200c}$.
  We subtract the $\Lambda$CDM two-halo term (Sec.~\ref{sec:method:twohalo}).
  We extrapolate the gas density profiles beyond $R_{\mathrm{max}}^X$ assuming a $1/r^4$ tail (left) and assuming the beta profile fits remain valid even at large radii where they were not well constrained by observations (right).
  Gray diamonds indicate the two clusters with the smallest $R_{\mathrm{max}}^X$ whose baryonic masses are likely to be underestimated with the $1/r^4$ extrapolation.
 }
\label{fig:fb}
\end{figure*}

\begin{figure*}
\begin{center}
\includegraphics[width=2\columnwidth]{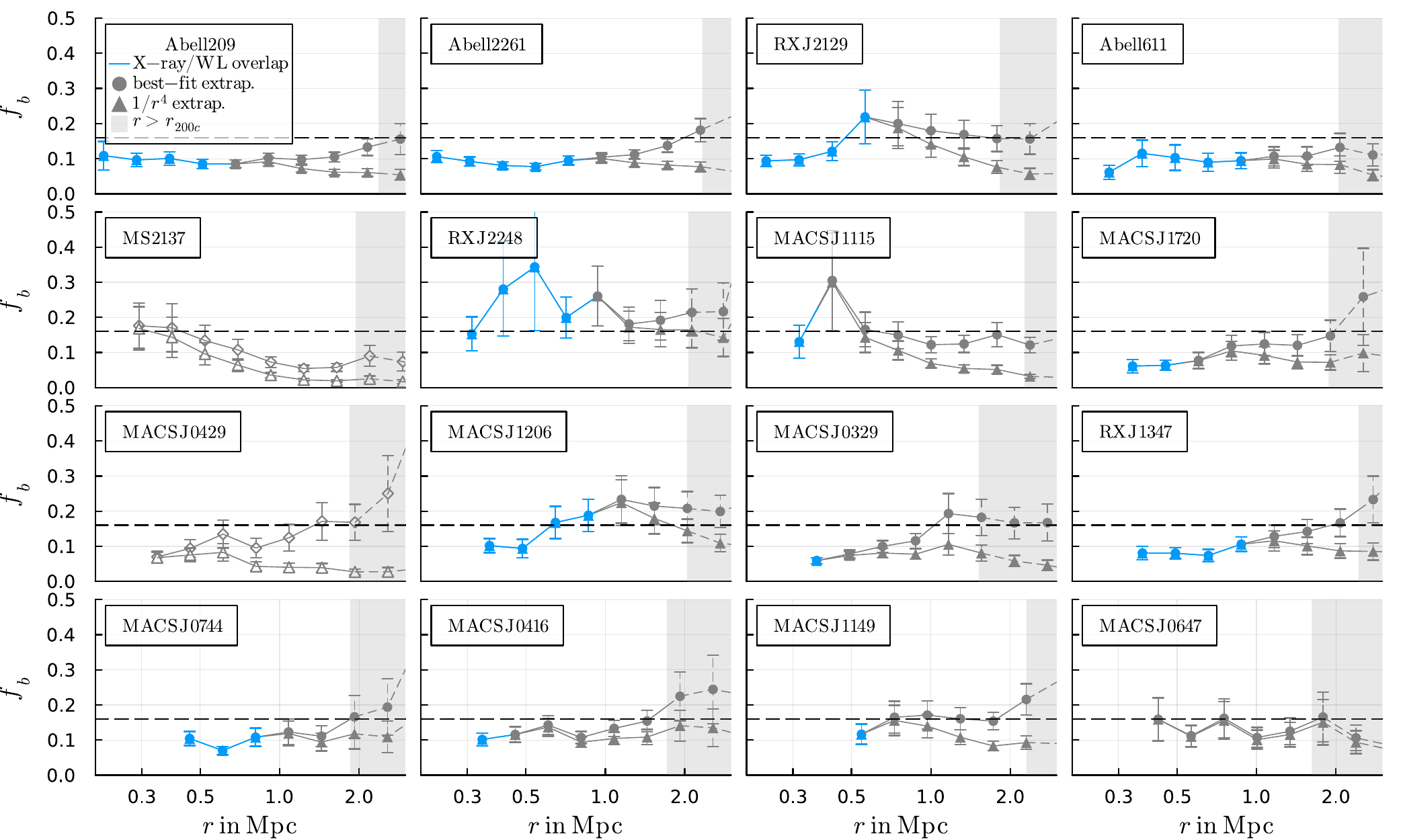}
\end{center}
\caption{%
 Same as Fig.~\ref{fig:fb} but separately for each cluster.
 Error bars indicate the statistical uncertainties.
 For each cluster, we show results assuming a $1/r^4$ gas density tail beyond $R_{\mathrm{max}}^X$ (smaller values of $f_b$, as in the left panel of Fig.~\ref{fig:fb}) and assuming that the best-fit beta profiles remain valid even beyond $R_{\mathrm{max}}^X$ (larger values of $f_b$, as in the right panel of Fig.~\ref{fig:fb}).
}
\label{fig:fb-grid}
\end{figure*}

Figure~\ref{fig:rho-grid} also shows the NFW density profiles from \citet{Umetsu2014}.
These were obtained by fitting the weak-lensing convergence profile plus the average convergence in the cluster center and match our non-parametric density profiles reasonably well.
As a further cross-check, we have fit NFW profiles to both our non-parametric mass and density profiles, finding results consistent with expectations, see Appendix~\ref{sec:appendix:nfwfit}.

\subsection{Cosmic baryon fraction}
\label{sec:results:fb}

The baryon fraction $\Omega_b/\Omega_m \approx 0.16$ \citep{PlanckCollaboration2018} plays an important role in cosmology.
One may expect that this cosmic baryon fraction is consistent with the ratio $f_b \equiv M_b / (M_b + M_{\mathrm{DM}})$ in a gravitationally collapsed structure with dark matter mass $M_{\mathrm{DM}}$ \citep[e.g.,][]{Planelles2013,Angelinelli2023,Rasia2025}.
This expectation is not realized in galaxies, where a much lower baryon fraction is detected.
In contrast, measurements of galaxy clusters at large radii seem to detect most of the expected baryons \citep[e.g.,][]{McGaugh2010,Wicker2023,Mantz2022}.

Figures~\ref{fig:fb} and \ref{fig:fb-grid} show the baryon fraction implied by our analysis as a function of radius.
The two-halo term is subtracted (Sec.~\ref{sec:method:twohalo}).
Our estimate of $f_b$ is most reliable where X-ray and weak-lensing observations overlap (blue lines and symbols).
In this radial range, most clusters in our sample have an $f_b$ well below the cosmic $0.16$.

At larger radii (gray lines and symbols), our results depend strongly on how we extrapolate the gas densities beyond the radius $R_{\mathrm{max}}^X$ where the beta-profile fits from \citet{Famaey2024} were well constrained by X-ray observations.
If we assume a $1/r^4$ density tail (see Sec.~\ref{sec:data:Mb}), most clusters remain well below the cosmic baryon fraction even at large radii.
This remains true even if we ignore the two clusters where $R_{\mathrm{max}}^X$ is likely too small to make a reliable estimate of the gas mass at large radii (diamond symbols).
This is consistent with the expectations for a MOND isothermal sphere, which is one motivation behind considering a $1/r^4$ tail.

On the other hand, if we extrapolate the gas densities by taking the beta profile fits at face value even at large radii, where they were not well constrained by observations, $f_b$ tends to increase with radius, with several clusters getting close to the cosmic baryon fraction around $r_{200c}$.
That matches expectations from $\Lambda$CDM, but we caution that the beta profiles from \citet{Famaey2024} may underestimate the steepness of the gas density profiles at such large radii (see Sec.~\ref{sec:data:Mb}).

Beyond $r_{200c}$ (dashed gray lines), the $f_b$ of some clusters grow significantly beyond the cosmic baryon fraction when taking the beta profile fits at face value.
However, the relative uncertainties become quite large at these extreme radii, so this effect is likely not significant.

\subsection{Stellar mass--halo mass relation}

\begin{figure}
\includegraphics[width=\columnwidth]{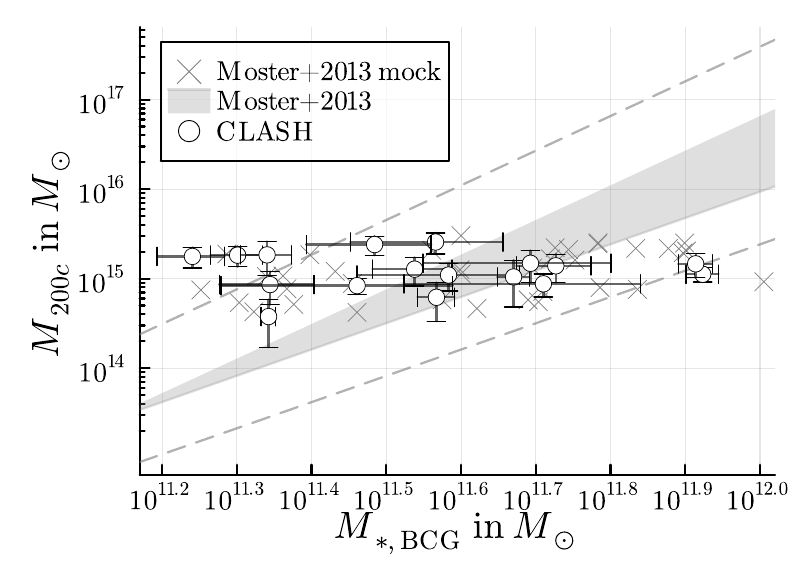}
\caption{%
 The SMHM relation implied by our non-parametric mass profiles (white symbols).
 The shaded gray region indicates the \citet{Moster2013} relation in the redshift range of our cluster sample.
 Dashed gray lines indicate $0.2 \,\mathrm{dex}$ scatter in the direction of $M_{\ast,M_{\mathrm{BCG}}}$ around that region.
 Gray crosses show simple mock data generated from the \citet{Moster2013} relation.
}
\label{fig:MBCG-vs-M200c}
\end{figure}

Figure~\ref{fig:MBCG-vs-M200c} shows the stellar--mass halo mass (SMHM) relation implied by our non-parametric mass profiles.
We exclude Abell 383 because, as discussed above, we cannot determine a value of $M_{200c}$.
We include RX J1532, whose BCG's stellar mass $M_{\ast,\mathrm{BCG}}$ is not provided by \citet{Famaey2024}, adopting its $M_{\ast,\mathrm{BCG}}$ directly from \citet{Burke2015}.

We find very little correlation between $M_{\ast,\mathrm{BCG}}$ and $M_{200c}$ \citep{Burke2015}.
This may seem unexpected because, in $\Lambda$CDM, a strong correlation between these two quantities should exist.
However, our cluster sample covers a relatively narrow range in total mass, so that what we see in Fig.~\ref{fig:MBCG-vs-M200c} may simply be the scatter in stellar mass at an essentially fixed $M_{200c}$.
This scatter is expected to be around $0.2\,\mathrm{dex}$.
Since the CLASH clusters cover a range of redshifts, the redshift evolution of the SMHM relation may add to the expected scatter.
With these considerations, the \citet{Moster2013} relation seems roughly consistent with our results (dashed gray lines in Fig.~\ref{fig:MBCG-vs-M200c}).
As a further check, we have generated simple mock data from the \citet{Moster2013} relation at $z=0.4$, close to the mean $z_l$ of our CLASH sample.
In particular, we generated $250$ equally-spaced $\log_{10} M_{200c}/M_\odot$ values between $10$ and $16.5$, calculated $M_{\ast,\mathrm{BCG}}$ according to the \citet{Moster2013} relation at $z=0.4$, then added, respectively, $0.1\,\mathrm{dex}$ and $0.2\,\mathrm{dex}$ noise to $M_{200c}$ and $M_{\ast,\mathrm{BCG}}$, and finally applied a cut $14.6 < \log_{10} M_{200c}/M_\odot < 15.5$. 
These mock data also seem consistent with our results (Fig.~\ref{fig:MBCG-vs-M200c}).

This is in contrast to massive spiral galaxies where the \citet{Moster2013} relation is in conflict with observations \citep[e.g.,][]{DiCintio2016}.
We also considered the SMHM relation from \citet{Kravtsov2018}, finding that it is significantly offset from our measurements towards higher $M_{\ast,\mathrm{BCG}}$.
However, the BCG stellar masses in \citet{Kravtsov2018} are defined to include stellar mass within many hundred $\mathrm{kpc}$, including contributions from the intracluster light, and are in fact extrapolated to infinity.
This contrasts with \citet{Burke2015} who measured $M_{\ast,\mathrm{BCG}}$ within $50\,\mathrm{kpc}$, which may explain the offset compared to \citet{Kravtsov2018}.

\begin{figure}
\begin{center}
\includegraphics[width=\columnwidth]{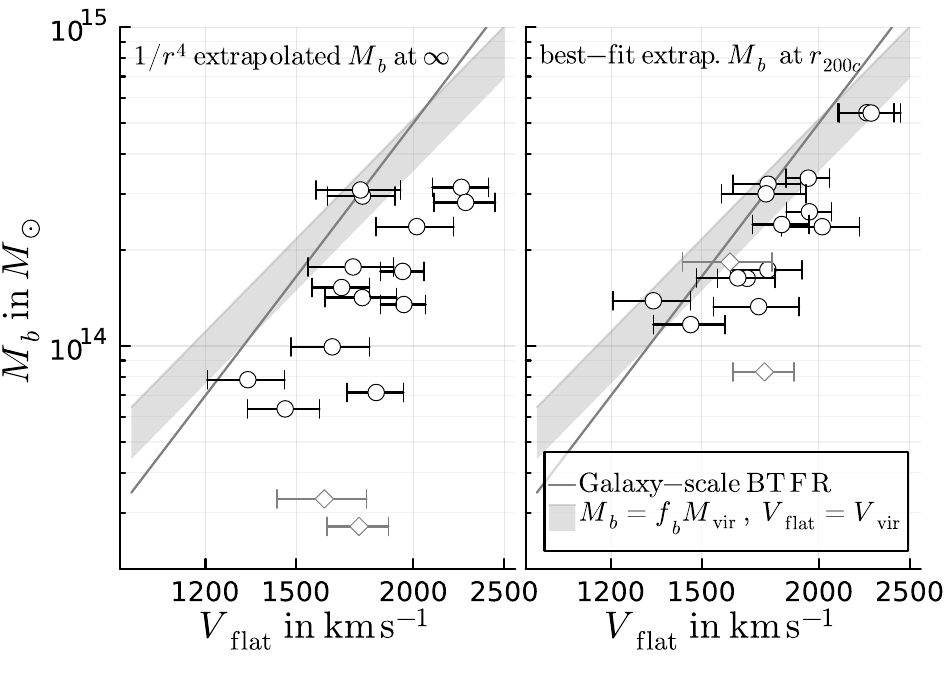}
\end{center}
\caption{%
 The BTFR implied by our non-parametric weak-lensing mass profiles and the baryonic mass estimates from \citet{Famaey2024}.
 Baryonic masses assume a $1/r^4$ gas density tail (left) or extrapolate the best-fit beta profiles out to $r_{200c}$ (right), see Sec.~\ref{sec:data:Mb}.
 Gray diamonds indicate the two clusters with the smallest radial range of X-ray observations, whose $M_b$ are likely significantly underestimated with the $1/r^4$ extrapolation. 
 The solid gray line is the galaxy-scale BTFR from \citet{Lelli2019}.
 The shaded band indicates a simple $\Lambda$CDM estimate with $V_{\mathrm{flat}} = \sqrt{G M_{\mathrm{vir}}/r_{\mathrm{vir}}}$, $M_{\mathrm{vir}} = f_b M_b$, and $f_b \approx 0.16$.
 The size of the band corresponds to the range of redshifts of the clusters shown.
}
\label{fig:BTFR-grid}
\end{figure}

\subsection{BTFR}
\label{sec:results:btfr}

\begin{figure*}
\begin{center}
\includegraphics[width=1.6\columnwidth]{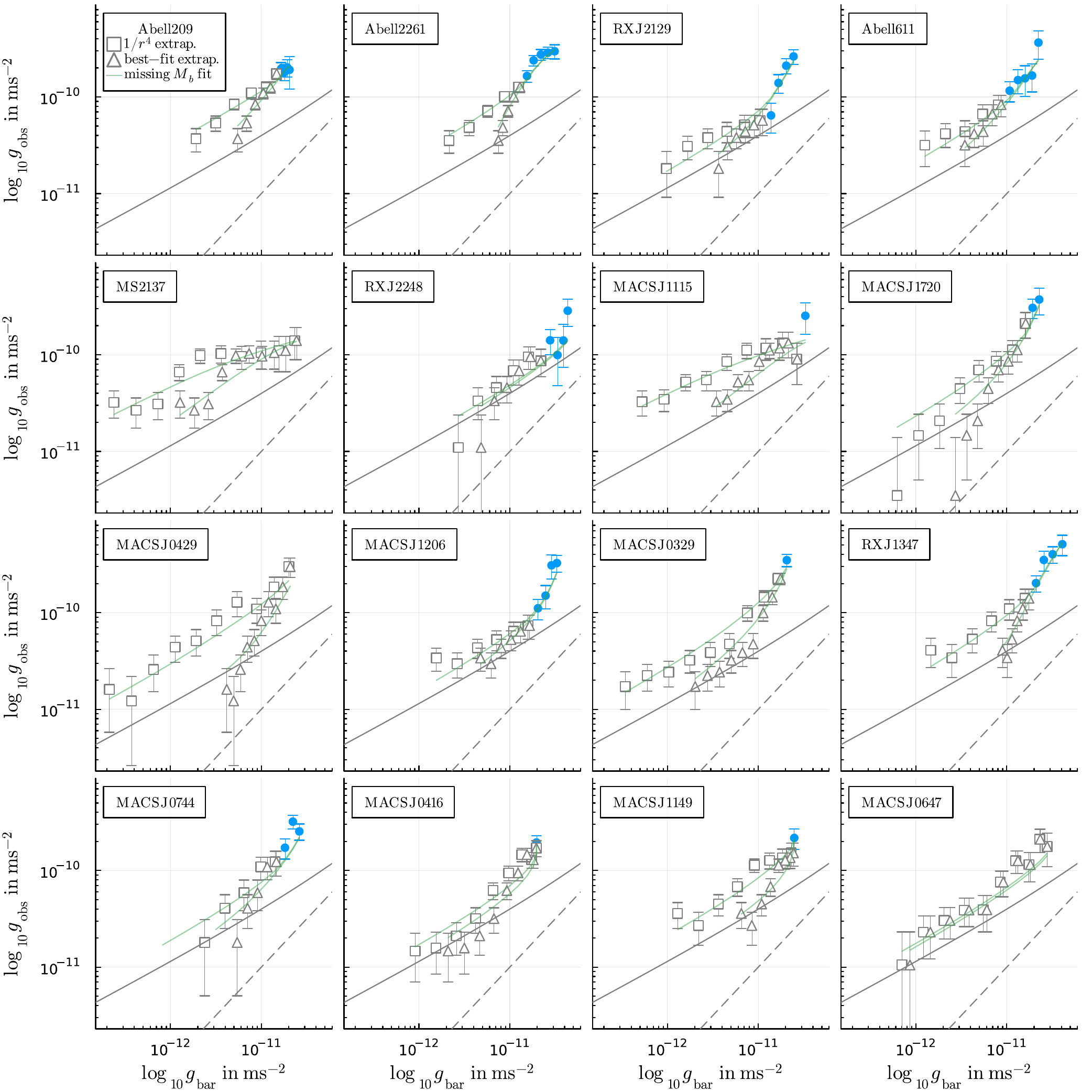}
\end{center}
\caption{%
 The RAR implied by our non-parametric weak-lensing mass profiles and the baryonic mass estimates from \citet{Famaey2024}.
 Blue circles indicate the radial range where X-ray and weak-lensing observations overlap.
 Beyond that range, we extrapolate the gas densities assuming a $1/r^4$ tail (gray squares) or assuming the best-fit beta profiles remain valid even at large radii where they were not well constrained by observations (gray triangles).
 We do not subtract the two-halo term (Sec.~\ref{sec:method:twohalo}).
 Two data points with negative $g_{\mathrm{obs}}$ for MACS J0744 (see Fig.~\ref{fig:M-grid}) are omitted.
 The dashed gray line indicates equality of $g_{\mathrm{obs}}$ and $g_{\mathrm{bar}}$, the solid gray line indicates the galaxy-scale RAR from \citet{Lelli2017b}.
 Assuming the galaxy-scale RAR holds for clusters, one can fit a missing baryonic component (not included in $g_{\mathrm{bar}}$) to $g_{\mathrm{obs}}$ (solid green lines, see Sec.~\ref{sec:results:missing}).
}
\label{fig:RAR-grid}
\end{figure*}

The BTFR relates the asymptotic flat circular velocity $V_{\mathrm{flat}}$ to the total baryonic mass $M_b$.
In Sec.~\ref{sec:results:vc}, we saw that the circular velocities of the CLASH clusters are approximately flat.
Thus, for simplicity we define $V_{\mathrm{flat}}$ as the weighted average of the circular velocities $V_c$ for
\begin{equation}
 750\,\mathrm{kpc} < r < 3\,\mathrm{Mpc} \,,
\end{equation}
with weights given by the inverse squares of the statistical uncertainties.
The circular velocities from Sec.~\ref{sec:results:vc} are not \emph{perfectly} flat, so our $V_{\mathrm{flat}}$ will change somewhat if we choose a different radial range.
We have verified that this effect is relatively minor and that other reasonable choices do not change our conclusions.

\begin{figure*}
\begin{center}
\includegraphics[width=2\columnwidth]{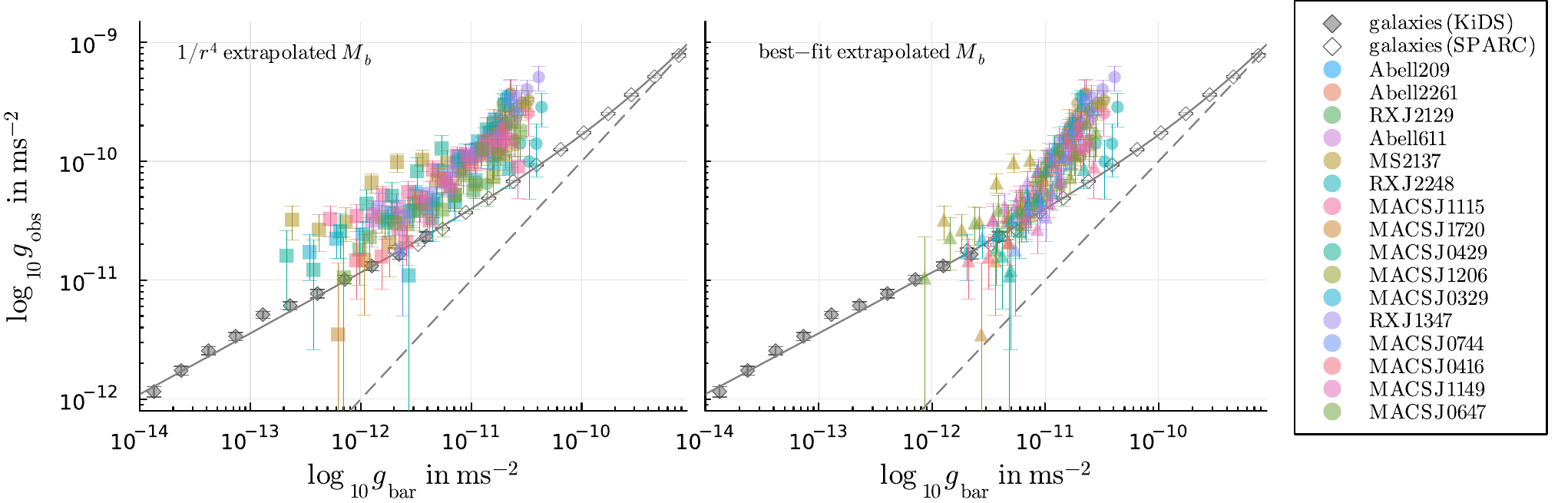}
\end{center}
\caption{%
 Same as Fig.~\ref{fig:RAR-grid}, but all clusters combined.
 Left: Extrapolating the gas densities with a $1/r^4$ tail.
 Right: Extrapolating the gas densities assuming the best-fit beta profiles.
 Symbols are as in Fig.~\ref{fig:RAR-grid}.
 Gray and white diamonds show the RAR of galaxies from SPARC \citep{Lelli2017b} and KiDS \citep{Mistele2023d,Lelli2024}, respectively.
 }
\label{fig:RAR-scatter}
\end{figure*}

Regarding the second ingredient of the BTFR, the total baryonic mass $M_b$, we face the issue that the gas masses $M_{\mathrm{gas}}(r)$ implied by the double beta fits from \citet{Famaey2024} are divergent.
This is not an issue for the $1/r^4$ extrapolation which integrates to a finite total mass (see Sec.~\ref{sec:data:Mb}), but we must deal with it when we extrapolate by taking the beta profile fits at face value.
Thus, in the latter case, we simply evaluate $M_b$ at the radius $r_{200c}$ implied by our non-parametric mass profiles (instead of at $r = \infty$).
This choice is somewhat ad-hoc but may have physical significance within $\Lambda$CDM, where the baryon fraction may be expected to be close to the cosmological one at the virial radius (see Sec.~\ref{sec:results:fb}).

The resulting BTFRs are shown in Fig.~\ref{fig:BTFR-grid}.
We do not subtract the two-halo term.
We have verified that it is not important.
How we estimate $M_b$ is important, however.
Extrapolating with a $1/r^4$ tail leads to a significant offset from the \citet{Lelli2019} galaxy-scale relation (solid line in Fig.~\ref{fig:BTFR-grid}), while evaluating the double beta fits at $r_{200c}$ leads to a comparably small offset.

In either case, our results are consistent with existing literature finding that galaxy clusters may follow a parallel relation compared to galaxies \citep{Sanders2003,McGaugh2015}.
With the $1/r^4$ extrapolation, the two clusters with the lowest $M_b$ clearly do not follow a parallel relation, but this is expected since their $M_{\mathrm{gas}}$ is likely significantly underestimated (Sec.~\ref{sec:data:Mb}).

Figure~\ref{fig:BTFR-grid} also shows a simple $\Lambda$CDM estimate based on identifying $(V_{\mathrm{flat}}, M_b)$ with $(\sqrt{G M_{\mathrm{vir}}/r_{\mathrm{vir}}}, f_b M_{\mathrm{vir}})$ where $f_b \approx 0.16$  is the cosmic baryon fraction.
This corresponds to a slope of 3 \citep{McGaugh2012} which is notably different from the galaxy-scale BTFR which has a slope of about 4.
A slope of about 4 seems to be a somewhat better representation of the trend in our results.

\subsection{RAR}
\label{sec:results:rar}

The RAR relates the Newtonian acceleration $g_{\mathrm{bar}}$ due to the baryonic mass at a given radius, $G_N M_b(r)/r^2$, to the total acceleration $g_{\mathrm{obs}}$, $G_N M(r)/r^2$, at that radius.
Unlike the BTFR, this relation can be evaluated locally at each radius (assuming spherical symmetry).
Thus, in the radial range where X-ray and weak-lensing observations overlap, we can measure the RAR without having to worry about how to extrapolate the gas profiles.

The RAR implied by our baryonic and total mass estimates of each galaxy cluster are shown in Fig.~\ref{fig:RAR-grid}.
Blue symbols indicate the range where X-ray and weak-lensing observations overlap.
Outside this range, we separately show extrapolations of $M_b$ assuming a $1/r^4$ gas density tail and assuming that the best-fit double beta profiles continue to be valid (Sec.~\ref{sec:data:Mb}).
We do not subtract the two-halo term.
Doing so does not significantly change the results.
Figure~\ref{fig:RAR-scatter} shows the data of all galaxy clusters combined.

The radial range where X-ray and weak-lensing observations overlap is quite narrow for many clusters and non-existent for 3 clusters.
Thus, we cannot draw strong conclusions about the galaxy cluster RAR at large radii or, equivalently, small accelerations.
We can, however, see that the galaxy-scale RAR seems to underpredict the $g_{\mathrm{obs}}$ of clusters already at relatively small radii, i.e. at relatively large $g_{\mathrm{bar}}$.
For a few clusters, this may be because the weak-lensing measurements of $g_{\mathrm{obs}}$ are not trustworthy at small radii (e.g. RX J2129, see Sec.~\ref{sec:results:vc}).
Another hypothesis \citep[e.g.,][]{Milgrom2008,Kelleher2024,Li2023,Famaey2024} is that there is an additional baryonic mass component not captured by our estimate of $g_{\mathrm{bar}}$.
That the galaxy-scale RAR underpredicts $g_{\mathrm{obs}}$ already at small radii means that, if it exists, this missing mass is not (only) to be found at large radii.
We discuss this more in Sec.~\ref{sec:results:missing}.

\subsection{Missing mass to recover galaxy-scale RAR}
\label{sec:results:missing}

In Sec.~\ref{sec:results:rar}, we saw that the galaxy-scale RAR underpredicts the $g_{\mathrm{obs}}$ observed in clusters.
Assuming that the RAR is a universal relation, this may be a sign that there is a missing baryonic mass component \citep{Eckert2022,Kelleher2024,Famaey2024} such as undetected, compact clouds of cold gas \citep{Milgrom2008}.
To test this hypothesis, \citet{Kelleher2024} have fit the observed $g_{\mathrm{obs}}$ in clusters by adding a missing $M_b$ component and assuming that the galaxy-scale RAR holds.
They assumed a missing mass density proportional to $(M_{\mathrm{mm,tot}}/r_s^3) (1+r/r_s)^{-4}$ and found that such a profile could reasonably explain the observed $g_{\mathrm{obs}}$ within about $1\,\mathrm{Mpc}$, where their data is likely not affected by hydrostatic bias.
The same missing mass profile was also considered by \citet{Famaey2024} among a variety of other profiles.
We here repeat the fitting procedure of \citet{Kelleher2024} using our measurements.

\begin{deluxetable*}{l|cccc|cccc}
\tablecaption{Fit parameters of the missing $M_b$ profiles following \citet{Kelleher2024}.}
\tablehead{
\colhead{} &
\multicolumn{4}{c}{$1/r^4$ extrapolated $M_b$} &
\multicolumn{4}{c}{best-fit extrapolated $M_b$}
\\
\colhead{Name} &
\colhead{$\log_{10} M_{\mathrm{mm,tot}}$} &
\colhead{$\log_{10} \rho_s$} &
\colhead{$\log_{10} r_s$} &
\colhead{$\log_{10} \Upsilon_b$} &
\colhead{$\log_{10} M_{\mathrm{mm,tot}}$} &
\colhead{$\log_{10} \rho_s$} &
\colhead{$\log_{10} r_s$} &
\colhead{$\log_{10} \Upsilon_b$}
\\
\colhead{} &
\colhead{$M_\odot$} &
\colhead{$M_\odot\,\mathrm{Mpc}^{-3}$} &
\colhead{$\mathrm{kpc}$} &
\colhead{} &
\colhead{$M_\odot$} &
\colhead{$M_\odot\,\mathrm{Mpc}^{-3}$} &
\colhead{$\mathrm{kpc}$} &
\colhead{}
}
\startdata
\input{./plots/table-missing-Mb-fits.tex}\enddata
\tablecomments{%
The listed values are the 16th, 50th, and 84th percentiles.
Values of $\rho_s = 3 M_{\mathrm{mm,tot}}/(4 \pi r_s^3)$ are provided for easier comparison to Table 1 of \citet{Famaey2024}.
}
\label{tab:missingMb}
\end{deluxetable*}

The fit has three free parameters, $M_{\mathrm{mm,tot}}$, $r_s$, as well as a parameter $\Upsilon_b$ that scales the observed baryonic mass, akin to a mass-to-light ratio.
We assume flat priors for $12 < \log_{10} M_{\mathrm{mm,tot}}/M_\odot < 16$ and $1 < \log_{10} r_s/\mathrm{kpc} < 4$ as well as a log-normal prior around $\Upsilon_b = 1$ with $0.1\,\mathrm{dex}$ scatter.
We do not subtract the two-halo term.
Subtracting the two-halo term does not significantly change the results.
For the RAR acceleration scale, we adopt $a_0 = 1.2 \cdot 10^{-10}\,\mathrm{m/s}^2$ \citep{Lelli2017b}.
The resulting best fits are shown as solid green lines in Fig.~\ref{fig:RAR-grid} and Table~\ref{tab:missingMb}.

Overall, we find that a missing baryonic component can reasonably fit our observations.
We find best-fit parameters roughly consistent with those from \citet{Kelleher2024} and \citet{Famaey2024}.
Finding a reasonable missing $M_b$ profile fails only when the observed galaxy-cluster RAR in Fig.~\ref{fig:RAR-grid} dips below the solid gray line, i.e. below the galaxy-scale RAR.
This is because, assuming the galaxy-scale RAR holds, such dips require negative missing mass densities.
This mostly only happens at the very largest radii, i.e. smallest $g_{\mathrm{bar}}$, and when extrapolating the gas densities by taking the beta profile fits from \citet{Famaey2024} at face value.
No significant negative mass densities are required when extrapolating the gas density with a $1/r^4$ tail, corresponding to a MOND isothermal sphere \citep{Milgrom1984}.
We also note that neighboring data points are positively correlated (see Fig.~\ref{fig:a209-correlation}) so that the visual impression of neighboring data points fluctuating down together may be misleading.
In any case, previous analyses have also seen this phenomenon.
At such large radii, however, analyses based on gas thermodynamics may suffer from hydrostatic bias \citep{Kelleher2024}, $g_{\mathrm{bar}}$ may not be well constrained \citep{Famaey2024}, and, depending on the theoretical framework, an external field effect may be important \citep{Kelleher2024}.

As an alternative approach, assuming that the galaxy-scale RAR holds universally, we can directly calculate and study the (non-parametric) missing mass profiles $M_b^{\mathrm{miss}}(r)$ implied by our measurements of $g_{\mathrm{obs}}$ and $g_{\mathrm{bar}}$.
We do so in Appendix~\ref{sec:appendix:missing}, finding similar results as above.

\section{Discussion}
\label{sec:discussion}

\subsection{Concentration and sparsity}

Our non-parametric results suggest a shift of perspective:
Away from first performing a parametric fit and then calculating quantities of interest in terms of the fit results, towards directly inferring quantities of interest from the non-parametric profiles $M(r)$ or $\rho(r)$, similar to how we measured the total mass $M_{200c}$ in Sec.~\ref{sec:results:vc} or the baryon fraction $M_b(r)/M(r)$ in Sec.~\ref{sec:results:fb}.
Another quantity of interest is the concentration $c$, which captures information about the shape of the mass profile.
Traditionally, definitions of $c$ are tied to specific parametric profiles such as the NFW profile.
A concentration can, however, also be defined in a non-parametric way.

One example is the ratio $c_{0.1} \equiv r_{200c}/r_{0.1}$ where $r_{0.1}$ is the radius that encloses $10\%$ of the total mass \citep{Yasin2023}.
Indeed, in principle, we can directly measure $c_{0.1}$ using our non-parametric mass profiles.
Unfortunately, just like parametrically defined concentrations (Appendix~\ref{sec:appendix:nfwfit}), $c_{0.1}$ is challenging to measure in practice without extending the radial range covered by weak-lensing observations \citep[for example using strong lensing,][]{Merten2015,Umetsu2016,Umetsu2025}.
In our current sample, only 4 clusters reach sufficiently small radii to determine $c_{0.1}$.

Another example is the sparsity, defined as $M_{200c}/M_{\Delta c}$ with, for example, $\Delta = 500$ or $\Delta = 1000$ \citep{Balmes2014}.
Initial results indicate that our non-parametric mass profiles tend to imply smaller sparsities than the NFW fits from \citet{Umetsu2014}, consistent with the results from \citet{Balmes2014}.
A proper statistical analysis as well as a comparison between different parametric and non-parametric concentration measures are left for future work.

\subsection{Density profiles}
\label{sec:discussion:density}

In Sec.~\ref{sec:results:density}, we saw that reconstructing non-parametric density profiles is viable but noisy with current data.
One way to ameliorate this may be to stack many clusters.
This is especially relevant given the large cluster samples with weak-lensing data that are expected to become available from instruments such as Euclid \citep{EuclidCollaboration2024c}, Roman \citep{Spergel2015}, or Rubin \citep{LSSTScienceCollaboration2009}.
This may enable model-independent constraints on quantities such as the splashback radius \citep[e.g.][]{Diemer2014,More2015b,More2016}, allowing to distinguish between different models of dark matter and modified gravity \citep{Adhikari2018}.

For example, in the context of MOND, one might expect a much weaker splashback signal than in $\Lambda$CDM.
Indeed, the splashback feature is a steep drop in density corresponding to accreting collisionless matter reaching its apocenter after first infall.
In $\Lambda$CDM, a splashback feature exists in both the dark matter and galaxy density profiles since both are collisionless.
Since dark matter is the dominant contribution to the gravitational potential, a splashback feature is then visible also in the weak-lensing signal.
In contrast, in MOND-like theories, the dominant contribution to the gravitational potential is due to the collisional intracluster medium, with only a sub-dominant contribution from the collisionless galaxies.
As a result, one may expect to see a weaker splashback feature in the weak-lensing signal.

When testing individual theoretical models, it may be easiest to follow $\Lambda$CDM-based analyses \citep[e.g.,][]{More2016} and fit a parametric density profile that is chosen according to the specific model under consideration.
Nonetheless, non-parametric constraints are desirable since they are easier to interpret and to apply to many models at once.
The feasibility of such non-parametric constraints will be studied in future work.

\subsection{Baryonic mass measurements}

Compared to our total mass measurements from weak lensing, the baryonic masses are relatively poorly constrained at large radii.
This limits our results regarding baryon fractions and regarding scaling relations such as the BTFR and RAR.
Better knowledge of the hot gas density profiles would enable stronger constraints on cosmology, dark matter, and modified gravity.
Indeed, as discussed in Sec.~\ref{sec:data:Mb}, results from the X-COP project suggest that, while our simple $1/r^4$ extrapolation is perhaps too steep, the beta profile fits from \citet{Famaey2024} may not be steep enough at large radii.
Currently, it is unknown where our cluster sample falls between these two extremes.

It would therefore be very useful to combine our non-parametric weak-lensing measurements with improved gas density profiles from follow-up X-ray observations.
Another possibility would be to stack the gas densities of a large sample of galaxy clusters, for example from eROSITA \citep{Bulbul2024,Kluge2024}, to increase the signal-to-noise ratio at large radii.
These could then be combined with stacked weak-lensing observations of the same sample from, for example, Euclid.

\subsection{Cluster cosmology}

Cluster cosmology aims to constrain cosmological parameters such as $\Omega_m$ and $\sigma_8$ using the high-mass end of the halo mass function.
This requires unbiased mass measurements.
We expect that the non-parametric methods discussed here will be useful in minimizing the impact of a number of important biases that affect such measurements.
In particular, many cluster cosmology analyses are based on measurements of $M_{200c}$.
Non-parametric measurements of $M_{200c}$ using our method will (see Sec.~\ref{sec:method:triaxial} and \citealt{Mistele2024b}): 1) Not be thrown off by baryonic effects that may change the shape of the mass profile, 2) not be thrown off by complex, non-symmetric mass distributions at the centers of clusters (e.g., late-stage mergers), 3) not be thrown off, on average, by triaxiality, 4) be less affected by miscentering (and there is an efficient way to correct for residual miscentering effects), and 5) apply directly to models beyond $\Lambda$CDM such as models based on ultra-light dark matter that may change the shape of the halo profile.
In addition, our mass reconstruction runs fast, taking only a few milliseconds per cluster to run.

The caveat is that, by making fewer assumptions, the statistical uncertainties become larger.
However, as shown in Sec.~\ref{sec:results:vc}, that is only a moderate effect in practice.
Also, with large cluster samples from upcoming and future surveys statistical uncertainties will no longer be the main limiting factor.
In fact, a trade-off in favor of significantly reduced systematic uncertainties, i.e. biases, at the cost of somewhat larger statistical error bars may be required in order to fully exploit the potential of instruments such as Euclid, Roman, or Rubin.

\section{Conclusion}
\label{sec:conclusion}

We have applied a new weak-lensing deprojection method to the CLASH sample of galaxy clusters.
We have inferred non-parametric mass and density profiles, which we have studied by themselves as well as in relation to the baryonic mass components.
We find:
(1) The implied circular velocities are approximately flat.
(2) The radially resolved baryonic mass fractions vary significantly from cluster to cluster and depend strongly on how we extrapolate the X-ray gas profiles at large radii, so it is unclear whether the CLASH clusters reach the cosmic baryon fraction expected in $\Lambda$CDM.
(3) The non-parametric masses are consistent with the $\Lambda$CDM SMHM relation.
(4) The CLASH clusters deviate from the BTFR and the RAR defined by galaxies, but the offset depends strongly on how we extrapolate the gas masses.
Contrary to some previous results based on hydrostatic equilibrium, we find that galaxy clusters may fall on the same BTFR and RAR as galaxies if one adds a suitable positive baryonic mass component.

Several of these results are limited by the baryonic masses being relatively poorly constrained at large radii.
Improving on this will unlock stronger constraints on cosmology, dark matter, and modified gravity.

\section*{Acknowledgements}
We thank Amel Durakovic, Paolo Tozzi, Pengfei Li, and Konstantin Haubner for helpful discussions.
This work was supported by the DFG (German Research Foundation) – 514562826.

Our non-parametric mass and density profiles, including correlation matrices, as well as the flat circular velocities implied by the mass profiles are available on Zenodo, 
\url{https://dx.doi.org/10.5281/zenodo.15476959}.

\bibliographystyle{mnras}
\bibliography{cluster-lensing-rar}

\begin{appendix}

\section{Two-halo term in $\Lambda$CDM}
\label{sec:appendix:twohalo}

\subsection{Subtraction procedure}

Here, we describe the details of the two-halo subtraction procedure from Sec.~\ref{sec:method:twohalo}.
The two-halo term becomes important only at projected radii $R$ that are sufficiently large for $\Sigma/\Sigma_{\mathrm{crit}}$ to be small.
As is clear from Eq.~\eqref{eq:shear_ESD_relation}, the distinction between $\Delta \Sigma$ and $G_+$ is then unimportant and we have to a good approximation $\Delta \Sigma = G_+$.
For our two-halo subtraction procedure, it then suffices to consider only the second step Eq.~\eqref{eq:M_from_ESD} of our deprojection procedure, i.e. the step that converts $\Delta \Sigma$ to the mass $M$.
The first step that converts $G_+$ to $\Delta \Sigma$ is unimportant.

The excess surface density we observe, $\Delta \Sigma_{\mathrm{obs}}$ contains contributions both from the galaxy cluster itself, $\Delta \Sigma_\oneh$, and from the two-halo term, $\Delta \Sigma_\twoh$,
\begin{equation}
 \label{eq:ESD_obs_is_1h_plus_2h}
 \Delta \Sigma_{\mathrm{obs}} (R) = \Delta \Sigma_\oneh (R) + \Delta \Sigma_\twoh (R) \,.
\end{equation}
When we apply the deprojection formula Eq.~\eqref{eq:M_from_ESD} from Sec.~\ref{sec:method:deprojection} to the observed $\Delta \Sigma_{\mathrm{obs}}$, we infer a mass profile $M_{\mathrm{obs}}$ that likewise contains the desired contribution from the galaxy cluster itself, $M_\oneh$, and a two-halo contribution $M_\twoh$,
\begin{equation}
 \label{eq:Mobs_incl_M2h}
 M_{\mathrm{obs}}(r) = M_\oneh(r) + M_\twoh(r)\,,
\end{equation}
where
\begin{equation}
 \label{eq:M2h_from_ESD2h_naive}
 M_\twoh(r) \equiv 4 r^2 \int_0^{\pi/2} d \theta \, \Delta \Sigma_\twoh \left( \frac{r}{\sin \theta}\right) \,.
\end{equation}
We note that $M_\twoh$ is not the mass profile of any actual object.
It just quantifies by how much our deprojection technique overestimates the true galaxy cluster mass profile when applied to lensing data that contains contributions from a cluster's local environment.
In practice, there is a technical complication, not captured by Eq.~\eqref{eq:M2h_from_ESD2h_naive}, due to the fact that we extrapolate $G_+$ beyond the last measured data point at $R_{\mathrm{max}}$.
This requires a modification of Eq.~\eqref{eq:M2h_from_ESD2h_naive} that we discuss separately in Appendix~\ref{sec:appendix:twohalo:wellactually}.
Thus, in practice, we use Eq.~\eqref{eq:appendix:M2h_from_ESD2h_tailed} instead of Eq.~\eqref{eq:M2h_from_ESD2h_naive}.

A simple estimate of $\Delta \Sigma_\twoh$ within $\Lambda$CDM is \citep[e.g.][]{Guzik2001,Oguri2011,Covone2014}
\begin{equation}
 \label{eq:ESD2h}
 \Delta \Sigma_\twoh (R) = b \frac{\bar{\rho}_{m,0}}{2 \pi D(z_l)^2} \int_0^\infty d \ell \, \ell J_2\left(\frac{\ell R}{D(z_l)}\right) P_m(k_\ell; z_l) \,,
\end{equation}
where $J_2$ denotes the second Bessel function of the first kind, $b$ is the bias, $\bar{\rho}_{m,0}$ is the mean matter density at redshift $z=0$, and $P_m(k_\ell; z_l)$ is the linear matter power spectrum at $k_\ell = \ell/[(1+z) D(z_l)]$.
We calculate $P_m$ using CAMB and we adopt the bias from \citet{Tinker2010} which gives $b$ as a function of the mass $M_{\Delta m}$ and redshift $z_l$,
\begin{equation}
b = b(z_l, M_{\Delta m}) \,.
\end{equation}
Here, $M_{\Delta m}$ is the mass within the radius $r_{\Delta m}$ where the cluster's average mass density drops below $\Delta$ times the cosmological matter density $\bar{\rho}_m$ at the cluster redshift $z_l$.
Below, we use $M_{200c}$ instead of $M_{\Delta m}$.
To avoid a mismatch, we use a value of $\Delta$ that makes the two halo mass definitions equivalent, namely $\Delta = 200 \rho_{\mathrm{crit}}(z_l)/\bar{\rho}_m (z_l)$.

Our goal is to recover the galaxy cluster's mass profile $M_\oneh$ by subtracting the two-halo contribution $M_\twoh$ from $M_{\mathrm{obs}}$.
To this end, we use the independent estimate of the two-halo contribution Eq.~\eqref{eq:ESD2h}.
This estimate of $\Delta \Sigma_\twoh$ depends on the redshift and the total mass of the galaxy cluster.
It can be converted to $M_\twoh$ using Eq.~\eqref{eq:M2h_from_ESD2h_naive} (or, rather, Eq.~\eqref{eq:appendix:M2h_from_ESD2h_tailed}).
We denote this estimate of $M_\twoh$ by
\begin{equation}
 M_\twoh(r|M_{200c}) \,,
\end{equation}
which makes the dependence on the total mass explicit.
Given the observed $M_{\mathrm{obs}}$, we can determine $M_{200c}$ by numerically solving the following equation for $M_{200c}$,
\begin{equation}
 M_{\mathrm{obs}} (r_{200c}) = M_{200c} + M_\twoh(r_{200c}|M_{200c}) \,,
\end{equation}
where $M_{200c} = (4\pi/3) \, 200 \cdot \rho_{\mathrm{crit}} \, r_{200}^3$.
This equation follows from Eq.~\eqref{eq:Mobs_incl_M2h} by using our estimate $M_\twoh(r|M_{200c})$ for the two-halo contribution and by using that, by definition, $M_\oneh(r_{200c}) = M_{200c}$.
Having determined $M_{200c}$ in this way, we can extract the desired mass profile $M_\oneh(r)$ of the galaxy cluster by subtracting the two-halo contribution,
\begin{equation}
 M_\oneh(r) = M_{\mathrm{obs}}(r) - M_\twoh(r|M_{200c})\,.
\end{equation}

The above procedure can easily be adapted to work with $M_{500c}$ instead of $M_{200c}$.
We will use $M_{500c}$ instead of $M_{200c}$ for one galaxy cluster, Abell 383, where a value for $M_{200c}$ cannot be determined.

\subsection{Adjustments due to extrapolation}
\label{sec:appendix:twohalo:wellactually}

The observed $\Delta \Sigma$ contains contributions from both the 1-halo and 2-halo terms, see Eq.~\eqref{eq:ESD_obs_is_1h_plus_2h}.
Thus, applying our deprojection formula Eq.~\eqref{eq:M_from_ESD} gives
\begin{equation}
 M_{\mathrm{obs}}(r) = 4 r^2 \int_0^{\pi/2} d\theta \, \left(
    \Delta \Sigma_\oneh\left(\frac{r}{\sin \theta}\right)
   + \Delta \Sigma_\twoh\left(\frac{r}{\sin \theta}\right)
  \right)\,, \quad \mathrm{(not\;quite\;correct\;in\;practice)} \,.
\end{equation}
This is what Eq.~\eqref{eq:M2h_from_ESD2h_naive} is based on, but it is not quite what we actually infer in practice.
The reason is that, after the last measured data point at $R_{\mathrm{max}}$, we no longer use $\Delta \Sigma = \Delta \Sigma_\oneh + \Delta \Sigma_\twoh$ in the integrand.
Instead, we use a power law extrapolation, $G_+ \propto 1/R^n$ (see Sec.~\ref{sec:method:deprojection}).
At the large radii relevant for $R \geq R_{\mathrm{max}}$, this extrapolation also implies $\Delta \Sigma \propto 1/R^n$ to a good approximation.
Thus, what we actually infer is
\begin{equation}
 M_{\mathrm{obs}}(r)
 = 4r^2 \left[
  \int\displaylimits_{\theta_{\mathrm{min}}}^{\pi/2} d\theta \, \left(
     \Delta \Sigma_\oneh\left(\frac{r}{\sin \theta}\right)
    + \Delta \Sigma_\twoh\left(\frac{r}{\sin \theta}\right)
   \right)
  + (\Delta \Sigma_\oneh(R_{\mathrm{max}}) + \Delta \Sigma_\twoh(R_{\mathrm{max}})) \frac{R^n_{\mathrm{max}}}{r^n} \int\displaylimits_0^{\theta_{\mathrm{min}}} d\theta \sin^n \theta
  \right]
  \,,
\end{equation}
where $\theta_{\mathrm{min}} = \arcsin(r/R_{\mathrm{max}})$.
If our power law extrapolation correctly captures the behavior of $\Delta \Sigma_\oneh$, this is
\begin{equation}
 M^{\mathrm{obs}}(r)
 = M_\oneh(r) + 4r^2 \left[
  \int_{\theta_{\mathrm{min}}}^{\pi/2} d\theta \, \Delta \Sigma_\twoh\left(\frac{r}{\sin \theta}\right)
  + \Delta \Sigma_\twoh(R_{\mathrm{max}}) \frac{R^n_{\mathrm{max}}}{r^n} \int_0^{\theta_{\mathrm{min}}} d\theta \sin^n \theta
  \right]
  \,.
\end{equation}
Indeed, for the extrapolation, one should choose a power law decay $1/R^n$ that plausibly matches the behavior of the shear due to the galaxy cluster itself, i.e. due to the one-halo term (\emph{not} the total shear including the two-halo term).
We can now read off the correct expression to use for $M_\twoh(r|M_{200c})$ in our two-halo subtraction procedure,
\begin{equation}
 \label{eq:appendix:M2h_from_ESD2h_tailed}
 M_\twoh(r|M_{200c}) =  4r^2 \left[
  \int_{\theta_{\mathrm{min}}}^{\pi/2} d\theta \, \Delta \Sigma_\twoh\left(\frac{r}{\sin \theta}\right)
  + \Delta \Sigma_\twoh(R_{\mathrm{max}}) \frac{R^n_{\mathrm{max}}}{r^n} \int_0^{\theta_{\mathrm{min}}} d\theta \sin^n \theta
  \right] \,.
\end{equation}
This replaces Eq.~\eqref{eq:M2h_from_ESD2h_naive} and takes into account that we extrapolate beyond the last measured data point $R_{\mathrm{max}}$.
See Appendix~\ref{sec:appendix:nfwextrapolation} for how to adopt this procedure when extrapolating assuming an NFW profile instead of a power law.

\section{Uncertainties and covariances}
\label{sec:appendix:uncertainties}

As systematic uncertainties, we consider our choices of how to extrapolate and interpolate the shear profile $G_+$.
To estimate the effect of these choices, we first calculate mass profiles with opposite and extreme choices of how to extrapolate $G_+$ beyond the last data point.
In particular, we consider extrapolating $G_+$ assuming $1/R^2$ and $1/\sqrt{R}$ power law decays.
These correspond, respectively, to a decay as fast as for a point particle and to a decay significantly slower than for an SIS.
These cases likely bracket the true behavior of the cluster's shear.
At each radius $r$, we calculate the minimum and maximum mass achievable in this way.
Schematically,
\begin{align}
M(r)^{\mathrm{max}} &= \max_{n \in \{\frac12, 1, 2\}} \left. M(r)\right|^{\mathrm{extrapolate}\;1/R^n} \,,
\\
M(r)^{\mathrm{min}} &= \min_{n \in \{\frac12, 1, 2\}} \left. M(r)\right|^{\mathrm{extrapolate}\;1/R^n} \,.
\end{align}
To take into account systematic uncertainties from interpolation, we add (subtract) the difference between the mass profiles obtained using linear and quadratic interpolation to $M^{\mathrm{max}}$ (from $M^{\mathrm{min}}$).
Schematically,
\begin{equation}
 M^{\substack{\mathrm{max}\\ \mathrm{min}}} \to 
 M^{\substack{\mathrm{max}\\ \mathrm{min}}}  \pm \left| \left.M(r)\right|^{\mathrm{quadratic}} - \left.M(r)\right|^{\mathrm{linear}} \right| \,.
\end{equation}

For the statistical uncertainties and covariances, we use linear error propagation to convert uncertainties and covariances on $G_+$ into uncertainties and covariances on the inferred mass $M$.
Following \citet{Mistele2024b}, we implement this error propagation by writing differentiable Julia code and then using `ForwardDiff.jl` \citep{Revels2016} to calculate the required Jacobians.
This reduces linear error propagation to a simple matrix multiplication.

For the covariance matrix of $G_+ = \langle g_+ \rangle/\langle \Sigma_{\mathrm{crit}}^{-1} \rangle$, we take into account uncertainties in $\langle g_+ \rangle$ as well as uncertainties and covariances due to $\langle \Sigma_{\mathrm{crit},ls}^{-1} \rangle$ and due to the correlated LSS \citep{Hoekstra2003},
\begin{equation}
 C^G = C^{G}_g + C^G_{\mathrm{crit}} + C^G_{\mathrm{LSS}} \,.
\end{equation}
We assume that the measurement uncertainties on $\langle g_+ \rangle$ in different radial bins are uncorrelated,
\begin{equation}
 \left(C^G_g\right)_{i j} = \delta_{i j} \frac{\sigma_{g_+,i}^2}{\langle \Sigma_{\mathrm{crit}}^{-1} \rangle^2} \,,
\end{equation}
where $i$ and $j$ run over the radial bins and $\sigma_{g_+,i}$ denotes the uncertainty on $\langle g_+ \rangle$ in the $i$-th radial bin.
As discussed in Sec.~\ref{sec:method:deprojection}, we assume $\langle \Sigma_{\mathrm{crit},ls}^{-1} \rangle$ to be the same in all radial bins.
This induces correlations between the radial bins,
\begin{equation}
 \left(C^G_{\mathrm{crit}}\right)_{i j} = \langle g_{+,i} \rangle \langle g_{+, j} \rangle \frac{\sigma_{\mathrm{crit}}^2}{\langle \Sigma_{\mathrm{crit}}^{-1} \rangle^4} \,,
\end{equation}
where $\sigma_{\mathrm{crit}}$ is the uncertainty on $\langle \Sigma_{\mathrm{crit},ls}^{-1} \rangle$.
These formulas follow from the definition Eq.~\eqref{eq:G+def} of $G_+$ in terms of $g_+$ and $\Sigma_{\mathrm{crit}}$.
The LSS contribution is important only at relatively large radii and following \citet{Umetsu2020} we calculate it from the non-linear matter power spectrum produced by CAMB \citep{Lewis2002}.
For simplicity, we assume all source galaxies to be located in a single source plane with effective redshift $z_{s,\mathrm{eff}}$ when calculating $C^G_{\mathrm{LSS}}$ \citep[see also][]{Miyatake2019}.
We take $z_{s,\mathrm{eff}}$ to be the source redshift whose critical surface density $\Sigma_{\mathrm{crit}}$ is $\langle \Sigma_{\mathrm{crit}}^{-1} \rangle$.

In addition to $G_+$, our deprojection method from Sec.~\ref{sec:method:deprojection} also has a second input, namely $f_c = \langle \Sigma_{\mathrm{crit},ls}^{-2} \rangle/\langle \Sigma_{\mathrm{crit},ls}^{-1} \rangle$.
For simplicity and because \citet{Umetsu2014} do not readily provide these, we do not take into account uncertainties in $f_c$.
We expect that doing so would only have a minor effect since $f_c$ only enters as the prefactor of $\Sigma/\Sigma_{\mathrm{crit}}$ in the relation $G_+ = \Delta \Sigma/(1  - f_c \cdot \Sigma/\Sigma_{\mathrm{crit}})$.
Indeed, this term is unimportant at large radii and gives only moderate corrections at small radii.
We expect the same to hold for corrections to the uncertainties and covariances induced by $f_c$.
In addition, we expect the uncertainties on $f_c$ to be less than $10\%$.
This is smaller than the typical uncertainties on the shear $G_+$ at small radii, which would further reduce the importance of the $f_c$ uncertainties.

\begin{figure}
\begin{center}
 \includegraphics[width=.47\columnwidth]{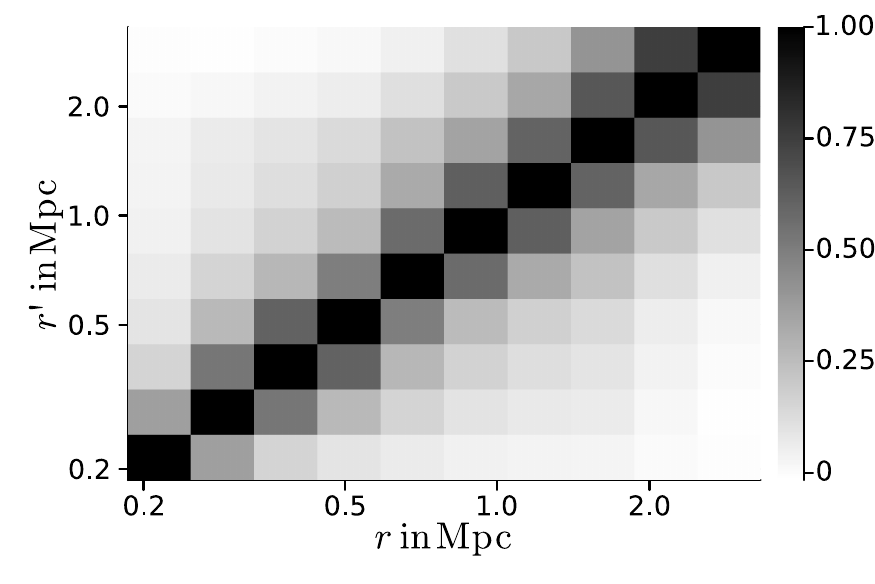}
\end{center}
\caption{The correlation matrix of the two-halo-subtracted masses inferred for Abell 209. The correlation matrix is defined in terms of the covariance matrix as $\mathrm{Cov}(M(r), M(r'))/\sigma_{M(r)} \sigma_{M(r')}$.}
\label{fig:a209-correlation}
\end{figure}

As mentioned above, we obtain the covariance matrix of the inferred masses $M(r)$ by linearly propagating the covariance matrix of $G_+ (R)$.
As a representative example, Fig.~\ref{fig:a209-correlation} shows the correlation matrix of the inferred, two-halo subtracted masses $M(r)$ for Abell 209.
We see that neighboring data points are correlated.
In part, these correlations are already present in the inputs to our method and are just propagated into the result.
But more importantly, the integrals Eq.~\eqref{eq:ESD_from_shear} and Eq.~\eqref{eq:M_from_ESD} mix different radii, producing additional correlations with the characteristic pattern previously found in \citet{Mistele2024}.

\section{Derivation of density reconstruction formula}
\label{sec:appendix:density}

We start with Eq.~\eqref{eq:M_from_ESD} which gives $M(r)$ in terms of $\Delta \Sigma(R)$,
\begin{equation}
 M(r) = 4 r^2 \int_0^{\pi/2} d\theta \, \Delta \Sigma\left(\frac{r}{\sin \theta}\right) \equiv 4 r^2 I(r) \,,
\end{equation}
where $I$ denotes the $\theta$ integral.
Using $\rho = M'(r) / 4 \pi r^2$, this gives
\begin{equation}
 \label{eq:appendix:densityreconstruction}
 \rho(r) = \frac{I'(r)}{\pi} + \frac{2 I(r)}{\pi r} \,.
\end{equation}
Our goal is to get rid of the derivative in the $I'(r)$ term.
The idea is to pull the derivative into the integrand of the $\theta$ integral, so that it acts on $\Delta \Sigma$, and then use integration by parts.
It will be useful to temporarily relax the upper integration boundary to $\pi/2-\epsilon$ and take the limit $\epsilon \to 0$ at the end of the calculation,
\begin{equation}
 I(r) = \lim_{\epsilon \to 0} I_\epsilon (r) \equiv \lim_{\epsilon \to 0} \int_0^{\frac\pi2 - \epsilon} d\theta \Delta \Sigma \left(\frac{r}{\sin \theta}\right) \,.
\end{equation}
We have
\begin{equation}
 I_\epsilon'(r)
 = \int_0^{\frac\pi2-\epsilon} d\theta \, \Delta \Sigma'\left(\frac{r}{\sin \theta}\right) \frac{1}{\sin \theta}
 = -\frac1r \int_0^{\frac\pi2-\epsilon} d\theta \, \frac{\sin \theta \tan \theta}{\sin \theta}\partial_\theta \left[ \Delta \Sigma\left(\frac{r}{\sin \theta}\right) \right] \,.
\end{equation}
After integrating by parts, this becomes
\begin{align}
 I_\epsilon'(r)
 &=-\frac1r \left[\tan \theta \, \Delta \Sigma\left(\frac{r}{\sin \theta}\right) \right]_{\theta=0}^{\theta=\frac\pi2-\epsilon}
  +\frac1r \int_0^{\frac\pi2-\epsilon} d\theta \, (\partial_\theta \tan \theta) \, \Delta \Sigma\left(\frac{r}{\sin \theta}\right)
 \\
 &= - \frac{\Delta \Sigma(r)}{\epsilon \, r} + \frac1r \int_0^{\frac\pi2-\epsilon} d\theta \, \Delta \Sigma\left(\frac{r}{\sin \theta}\right) \frac{1}{\cos^2 \theta} + O(\epsilon) \,.
\end{align}
The $1/\epsilon$ divergence is why it was useful to relax the upper integration boundary.
Importantly, a $1/\epsilon$ divergence exists not only in the boundary term but also in the remaining integral.
In fact, the divergence from the remaining integral precisely cancels the $1/\epsilon$ from the boundary term.
This must be the case because the expression we started with was finite in the limit $\epsilon \to 0$.
The factor of $\Delta \Sigma\left(\frac{r}{\sin \theta}\right)$ in the remaining integral is well-behaved at $\theta \to \pi/2$, i.e. at the upper integration boundary in the limit $\epsilon \to 0$.
The $1/\epsilon$ divergence is due only to the $1/\cos^2 \theta$ factor.
In fact,
\begin{equation}
 \int^{\frac\pi2-\epsilon}_0 \frac{d\theta}{\cos^2 \theta} = \frac1\epsilon + O(\epsilon) \,.
\end{equation}
We can therefore rewrite the $1/\epsilon$ from the boundary term as this $1/\cos^2 \theta$ integral and combine all integrals into one,
\begin{equation}
 I_\epsilon'(r) = \frac{1}{r} \int_0^{\frac\pi2 - \epsilon} d\theta \, \frac{\Delta \Sigma \left(\frac{r}{\sin \theta}\right) - \Delta \Sigma(r)}{\cos^2 \theta} + O(\epsilon) \,.
\end{equation}
The integrand of this integral is finite everywhere in the interval $(0, \pi/2)$.
This can be verified by expanding the integrand around $\pi/2$.
The limit $\epsilon \to 0$ can now be taken,
\begin{equation}
 I'(r) = \lim_{\epsilon \to 0} I'_\epsilon(r) = \frac1r \int_0^{\frac\pi2} d\theta \, \frac{\Delta \Sigma \left(\frac{r}{\sin \theta}\right) - \Delta \Sigma(r)}{\cos^2 \theta} \,.
\end{equation}
Plugging this result into Eq.~\eqref{eq:appendix:densityreconstruction} gives the desired result Eq.~\eqref{eq:densityreconstruction}.

\section{Line-of-sight average of inferred mass}
\label{sec:appendix:los-average}

As discussed in Sec.~\ref{sec:method:triaxial}, our deprojection formulas from Sec.~\ref{sec:method:deprojection} were derived assuming spherical symmetry.
Here we show that, nevertheless, if we average over all line-of-sight directions, our deprojection formulas Eq.~\eqref{eq:ESD_from_shear} and Eq.~\eqref{eq:M_from_ESD} produce the true mass $M_{\mathrm{true}}$ (see Eq.~\eqref{eq:Mtrue}) even for non-symmetric mass distributions.
This result holds given that two conditions are satisfied: (i) We restrict to the radial range where $\Sigma/\Sigma_{\mathrm{crit}}$ is negligible, and (ii) the source galaxy redshifts follow probability distributions that do not depend on the azimuth (but may depend on projected radius $R$).
The latter condition disallows a lopsidedness in the source galaxy population, but does allow a radial variation, for example due to obscuration towards the cluster center.

To show that this result holds, we first note that the true mass $M_{\mathrm{true}}(r)= \int_{|\vec{x}'| < r} d^3\vec{x}' \rho(\vec{x}')$ (Eq.~\eqref{eq:Mtrue}) can be written as a spherical integral over a spherical density $\bar{\rho} (r)$ obtained by averaging the 3D density $\rho(\vec{x})$ over the solid angle,
\begin{equation}
 \label{eq:los-average:Mtrue_rhobar}
 M_{\mathrm{true}} (r) = 4 \pi \int_0^r dr' r'^2 \bar{\rho}(r')
  \quad \mathrm{with}  \quad
 \bar{\rho}(|\vec{x}|) \equiv \frac{1}{4 \pi} \int d\Omega \, \rho(R(\Omega) \cdot \vec{x}) \,,
\end{equation}
where $R(\Omega)$ is a rotation matrix that implements rotation by $\Omega$.
The proof below then proceeds roughly as follows:
Due to the assumption that $\Sigma/\Sigma_{\mathrm{crit}} \ll 1$, the deprojection procedure from Sec.~\ref{sec:method:deprojection} becomes \emph{linear} in the density $\rho$ (see the denominator of the right-hand side in Eq.~\eqref{eq:shear_ESD_relation}).
As a result, averaging the inferred mass $M_{\mathrm{inferred}}$ over the line of sight becomes equivalent to running the deprojection procedure for a (fictitious) cluster that has a mass density $\bar{\rho} (r)$.
Since $\bar{\rho}$ is spherically symmetric, the deprojection procedure correctly infers the mass associated with $\bar{\rho} (r)$ which, according to Eq.~\eqref{eq:los-average:Mtrue_rhobar}, is just $M_{\mathrm{true}}$.
In the following, we work out these steps in detail.

The deprojection formulas from Sec.~\ref{sec:method:deprojection} are based on the observable $G_+ = \langle g_+ \rangle/\langle \Sigma_{\mathrm{crit}}^{-1} \rangle$.
As a first step, we will show that, in the radial range where $\Sigma/\Sigma_{\mathrm{crit}} \ll 1$, we have
\begin{equation}
 \label{eq:los-average:G+simplified}
 G_+ (R) = \Delta \Sigma_g (R) \quad \mathrm{with} \quad \Delta \Sigma_g (R) \equiv \frac{M_{2D}(R)}{\pi R^2} - \langle \Sigma \rangle (R)\,,
\end{equation}
where $\langle \Sigma \rangle$ is the azimuthally-averaged surface density and the subscript $g$ in $\Delta \Sigma_g$ indicates that this definition of $\Delta \Sigma$ applies more generally than the definition given in Eq.~\eqref{eq:ESDdef} which applies only in spherical symmetry.
Assuming spherical symmetry and $\Sigma/\Sigma_{\mathrm{crit}} \ll 1$, we can obtain the very similar result $G_+ = \Delta \Sigma$ from Eq.~\eqref{eq:shear_ESD_relation} (with $\Delta \Sigma$ defined by Eq.~\eqref{eq:ESDdef}).
The important difference is that Eq.~\eqref{eq:los-average:G+simplified} holds without any symmetry assumptions on the mass distribution.
To see this, we first consider the average $\langle g_+ \rangle(R)$ of the reduced tangential shear $g_+ = \gamma_+/(1 - \Sigma/\Sigma_{\mathrm{crit}}$) \citep{Bartelmann2001}.
The average $\langle \dots \rangle$ can be understood as, first, separately at each position $(R \cos \varphi, R \sin \varphi)$ averaging over the source redshifts $z_s$, and then averaging azimuthally over $\varphi$,
\begin{align}
 \langle g_+ \rangle (R)
 = \int \frac{d\varphi}{2 \pi} \int d z_s \, p(z_s|R, \varphi) \, g_+(R, \varphi)
  = \int \frac{d\varphi}{2 \pi} \int d z_s \, p(z_s|R, \varphi) \, \gamma_+(R, \varphi) \,,
\end{align}
where we used the assumption $\Sigma/\Sigma_{\mathrm{crit}} \ll 1$, which implies that $g_+$ becomes $\gamma_+$.
Further, our assumption that the source redshifts are drawn from probability distributions that do not depend on the azimuth means that $p(z_s|R, \varphi)$ satisfies $p(z_s|R, \varphi) = p(z_s|R)$.
Thus, we can pull the $z_s$ integral including the factor of $p(z_s|R)$ outside the $\varphi$ integral,
\begin{equation}
 \langle g_+ \rangle (R)
 = \int d z_s \, p(z_s|R) \Sigma_{\mathrm{crit},ls}^{-1} \cdot \int d\varphi \, \gamma_+(R, \varphi) \Sigma_{\mathrm{crit},ls} \,,
\end{equation}
where we also judiciously introduced factors of $\Sigma_{\mathrm{crit},ls}$.
Importantly, the azimuthal average of $\gamma_+ \Sigma_{\mathrm{crit}}$ gives $\Delta \Sigma_g$ without having to assume any symmetry of the mass distribution \citep{Kaiser1995,Bartelmann1995}.
Since $\Delta \Sigma_g$ is a property of only the lens, it is independent of the source redshifts $z_s$, and we can pull it outside the $z_s$ integral,
\begin{equation}
 \langle g_+ \rangle (R)
 = \Delta \Sigma_g (R) \cdot \int d z_s \, p(z_s|R) \Sigma_{\mathrm{crit},ls}^{-1}
 = \Delta \Sigma_g (R) \cdot  \langle \Sigma_{\mathrm{crit},ls}^{-1} \rangle \,.
\end{equation}
This is the desired result Eq.~\eqref{eq:los-average:G+simplified} after using the definition $G_+ = \langle g_+ \rangle/\langle \Sigma_{\mathrm{crit}}^{-1} \rangle$.

We note that, as discussed in \citet{Mistele2024b}, if individual redshift estimates for the source galaxies are available, the definition of $G_+$ can be changed to $G_+ = \langle \Sigma_{\mathrm{crit},ls} \, g_+ \rangle$.
In this case, the desired Eq.~\eqref{eq:los-average:G+simplified} follows even without having to assume that $p(z_s|R, \varphi)$ is independent of $\varphi$.

In any case, the input to our deprojection formulas Eq.~\eqref{eq:ESD_from_shear} and Eq.~\eqref{eq:M_from_ESD} from Sec.~\ref{sec:method:deprojection} are now $G_+ (R) = \Delta \Sigma_g (R)$ and $f_c$.
The first of these two deprojection formulas, Eq.~\eqref{eq:ESD_from_shear}, reduces to ``$\Delta \Sigma$'' $= \Delta \Sigma_g$ in the radial range where $\Sigma/\Sigma_{\mathrm{crit}}$ is negligible (here, ``$\Delta \Sigma$'' is understood as the result of evaluating the right-hand side of Eq.~\eqref{eq:ESD_from_shear}, which is to be inserted into Eq.~\eqref{eq:M_from_ESD}. Its formal definition Eq.~\eqref{eq:ESDdef} does not apply outside spherical symmetry).
This follows from Eq.~\eqref{eq:shear_ESD_relation} by using the fact that $\Sigma/\Sigma_{\mathrm{crit}} \ll 1$ implies $f_c \Sigma \ll 1$ for any reasonably-behaved source redshift distribution.
Thus, this first deprojection step is trivial in the radial range where $\Sigma/\Sigma_{\mathrm{crit}} \ll 1$.
The remaining second step of the deprojection procedure is Eq.~\eqref{eq:M_from_ESD} with $\Delta \Sigma_g$ as the integrand.
Specifically, the mass $M_{\mathrm{inferred}}$ we infer is
\begin{equation}
 M_{\mathrm{inferred}} (r)
 \equiv 4 r^2 \int_0^{\pi/2} d \theta \, \Delta \Sigma_g\left( \frac{r}{\sin \theta} \right)
 = 4 r^2 \int_0^{\pi/2} d \theta \, \left.\left(\frac{M_{2D}(R)}{\pi R^2} - \langle \Sigma \rangle (R) \right)\right|_{R = \frac{r}{\sin \theta}}
 \,. 
\end{equation}
Here, $M_{2D}(R)$ is the mass enclosed in the line-of-sight cylinder with radius $R$, which can be written as an integral over the azimuthally-averaged surface density $\langle \Sigma \rangle$, 
\begin{equation}
 \label{eq:los-averaged:Minferred1}
 M_{\mathrm{inferred}} (r)
 = 4 r^2 \int_0^{\pi/2} d \theta \, \left.\left(
  \frac{2}{R^2} \int_0^R dR' R' \langle \Sigma \rangle (R')
  - \langle \Sigma \rangle (R)
  \right)\right|_{R = \frac{r}{\sin \theta}}
 \,. 
\end{equation}
The right-hand side now depends only on the azimuthally-averaged surface density $\langle \Sigma \rangle$.

Consider averaging this $M_{\mathrm{inferred}}$ over all line-of-sight directions.
This amounts to calculating
\begin{equation}
 \label{eq:Minferred_los_average}
 \frac{1}{4\pi} \int d\Omega \, M_{\mathrm{inferred}}^{\Omega} (r) \,,
\end{equation}
where the integral over the solid angle $\Omega$ corresponds to the line-of-sight average and $M_{\mathrm{inferred}}^\Omega$ denotes the mass we infer when the underlying density is rotated by $\Omega$ with respect to the original mass distribution.
We refer to the density of the rotated mass distribution as $\rho^\Omega$,
\begin{equation}
 \rho^\Omega (\vec{x}) \equiv \rho( R(\Omega) \cdot \vec{x}) \,.
\end{equation}
Similarly, we denote the azimuthally-averaged surface density corresponding to $\rho^\Omega$ by $\langle \Sigma \rangle^\Omega$.
Concretely, $M_{\mathrm{inferred}}^\Omega$ is given by Eq.~\eqref{eq:los-averaged:Minferred1} with $\langle \Sigma \rangle$ replaced by $\langle \Sigma \rangle^\Omega$.
Let's explicitly write out this expression for $M_{\mathrm{inferred}}^\Omega$ in terms of $\rho^\Omega$,
\begin{equation}
 4 r^2 \int_0^{\pi/2} d\theta \int \frac{d\varphi}{2\pi} \int dz \left(
   \frac{2}{(r/\sin \theta)^2} \int_0^{\frac{r}{\sin \theta}} d R' R' \rho^\Omega(R' \cos \varphi, R' \sin \varphi, z)
   - \rho^\Omega\left(\frac{r}{\sin \theta} \cos \varphi, \frac{r}{\sin \theta} \sin \varphi, z\right)
 \right) \,.
\end{equation}
This is linear in $\rho^\Omega$ and depends on $\Omega$ only through $\rho^\Omega$.
Thus, when averaging $M_{\mathrm{inferred}}^\Omega$ over the line-of-sight directions as in Eq.~\eqref{eq:Minferred_los_average}, we can move the $\Omega$ integral past the other integrals and find
\begin{align}
 \frac{1}{4\pi} \int d\Omega \, M_{\mathrm{inferred}}^\Omega
  &= 4 r^2 \int_0^{\pi/2} d\theta \int dz \left.\left(
   \frac{2}{R^2} \int_0^{R} d R' R' \bar{\rho}\left(\sqrt{R'^2+z^2}\right)
   - \bar{\rho}\left(\sqrt{ R^2 + z^2 }\right)
  \right)\right|_{R = \frac{r}{\sin \theta}}
  \\
  &= 4 r^2 \int_0^{\pi/2} d\theta \left.\left(
   \frac{2}{R^2} \int_0^{R} d R' R' \bar{\Sigma}(R')
   - \bar{\Sigma}(R)
  \right)\right|_{R = \frac{r}{\sin \theta}} \,,
\end{align}
where $\bar{\rho}(r) = \frac{1}{4\pi} \int d\Omega \, \rho(R(\Omega) \cdot \vec{x})$ as in Eq.~\eqref{eq:los-average:Mtrue_rhobar}.
The integrand of the $\theta$ integral on the right-hand side is the excess surface density $\Delta \bar{\Sigma} (R)$ of a (fictitious) lens with density $\bar{\rho}$ evaluated at $R = r/\sin \theta$,
\begin{align}
 \frac{1}{4\pi} \int d\Omega \, M_{\mathrm{inferred}}^\Omega
  &= 4 r^2 \int_0^{\pi/2} d\theta \Delta \bar{\Sigma}\left(\frac{r}{\sin \theta}\right) \,.
\end{align}
Since that fictitious lens is spherically symmetric, the $\theta$ deprojection integral will infer the corresponding mass $\bar{M}$,
\begin{equation}
 \frac{1}{4\pi} \int d\Omega \, M_{\mathrm{inferred}}^\Omega = \bar{M}(r) \equiv 4 \pi \int_0^r dr' r'^2 \bar{\rho}(r') \,.
\end{equation}
According to Eq.~\eqref{eq:los-average:Mtrue_rhobar}, this is the same as $M_{\mathrm{true}}$, which is what was to be shown.

\section{NFW extrapolation}
\label{sec:appendix:nfwextrapolation}

As an alternative to extrapolating $G_+$ beyond the last measured data point by assuming a power law, we here consider extrapolation assuming an NFW profile.
In particular, we assume that, beyond the last measured data point at $R = R_{\mathrm{max}}$, the shear profile $G_+ = \langle g_+ \rangle / \langle \Sigma_{\mathrm{crit}}^{-1} \rangle$ is given by (see Eq.~\eqref{eq:shear_ESD_relation})
\begin{equation}
 \label{eq:nfwextrapolate}
 G_+ (R > R_{\mathrm{max}}) = \frac{\Delta \Sigma_{\mathrm{NFW}}(R|M_{200c}^{\mathrm{match}})}{1 - f_c \, \Sigma_{\mathrm{NFW}}(R|M_{200c}^{\mathrm{match}})} \,,
\end{equation}
where $\Delta \Sigma_{\mathrm{NFW}}(R|M_{200c})$ and $\Sigma_{\mathrm{NFW}}(R|M_{200c})$ denote, respectively, the excess surface density and surface density of an NFW halo with a given $M_{200c}$.
Explicit formulas for $\Delta \Sigma_{\mathrm{NFW}}$ and $\Sigma_{\mathrm{NFW}}$ are given, for example, in \citet{Umetsu2020}.
For simplicity, we here fix the concentration $c_{200c}$ by assuming the WMAP5 mass-concentration relation from \citet{Maccio2008}, so that the NFW profiles are fully specified by $M_{200c}$ alone.
Choosing a different mass-concentration relation does not significantly change our results.
When extrapolating $G_+$ using Eq.~\eqref{eq:nfwextrapolate}, we determine $M_{200c}$ by matching to the observed shear $G_+$ at the last data point at $R_{\mathrm{max}}$ and denote the result by $M_{200c}^{\mathrm{match}}$,
\begin{equation}
 \label{eq:M200cmatchdef}
 G_+(R=R_{\mathrm{max}}) = \frac{\Delta \Sigma_{\mathrm{NFW}}(R_{\mathrm{max}}|M_{200c}^{\mathrm{match}})}{1 - f_c \, \Sigma_{\mathrm{NFW}}(R_{\mathrm{max}}|M_{200c}^{\mathrm{match}})} \,.
\end{equation}
We note that $M_{200c}^{\mathrm{match}}$ does not necessarily coincide with the cluster's actual $M_{200c}$ as inferred  from the full, deprojected mass profile $M(r)$, hence the different notation.

The two-halo subtraction procedure described in Sec.~\ref{sec:method:twohalo} and Appendix~\ref{sec:appendix:twohalo} is mostly unchanged when extrapolating with an NFW profile.
We only need to replace Eq.~\eqref{eq:appendix:M2h_from_ESD2h_tailed} for the two-halo contribution to the inferred mass $M_\twoh$ with
\begin{multline}
 M_\twoh(r|M_{200c}) =  4r^2 \left[
  \int_{\theta_{\mathrm{min}}}^{\pi/2} d\theta \, \Delta \Sigma_\twoh\left(\frac{r}{\sin \theta}\right)
  \right.
  \\
  \left.
  + \int_0^{\theta_{\mathrm{min}}} d\theta \left(
   \Delta \Sigma_{\mathrm{NFW}}\left(\frac{r}{\sin \theta} \Big| M_{200c}^{\mathrm{match,obs}} \right) -
   \Delta \Sigma_{\mathrm{NFW}}\left(\frac{r}{\sin \theta} \Big| M_{200c}^{\mathrm{match,sub}} \right)
  \right)
  \right] \,.
\end{multline}
Here, $M_{200c}^{\mathrm{match,obs}}$ is determined by matching an NFW profile to the observed $G_+(R_{\mathrm{max}})$ as in Eq.~\eqref{eq:M200cmatchdef}, and $M_{200c}^{\mathrm{match,sub}}$ is obtained by matching an NFW profile to $G_+(R_{\mathrm{max}})$ minus the two-halo contribution, i.e. by matching an NFW profile to $G_+ - G_{+,\twoh}$ at $R = R_{\mathrm{max}}$.
We note that the two-halo contribution $G_{+,\twoh}$ depends on the actual $M_{200c}$ (as determined by the full, deprojected mass profile $M(r)$) through the bias factor $b(z_l, M_{200c}$) (see Appendix~\ref{sec:appendix:twohalo}) and, therefore, so does $M_{200c}^{\mathrm{match,sub}}$.

In practice, since it makes the code simpler and faster, we actually determine $M_{200c}^{\mathrm{match,sub}}$ from the equation
\begin{equation}
 \Delta \Sigma_{\mathrm{NFW}}(R_{\mathrm{max}}|M_{200c}^{\mathrm{match,obs}}) - \Delta \Sigma_\twoh(R_{\mathrm{max}}) = 
 \Delta \Sigma_{\mathrm{NFW}}(R_{\mathrm{max}}|M_{200c}^{\mathrm{match,sub}}) \,.
\end{equation}
If the two-halo term is negligibly small at $R_{\mathrm{max}}$, this gives $M_{200c}^{\mathrm{match,obs}} = M_{200c}^{\mathrm{match,sub}}$ which is the correct outcome in this case, i.e. it is what we would have obtained from matching an NFW profile to $G_+ - G_{+,\twoh}$ at $R_{\mathrm{max}}$.
When the two-halo term becomes non-negligible, we are very likely at sufficiently large radii for the difference between $G_+$ and $\Delta \Sigma$ to be unimportant (see Appendix~\ref{sec:appendix:twohalo}).
Thus, in this case, the outcome will again match the outcome of matching to $G_+ - G_{+,\twoh}$.
This justifies our simplified procedure for determining $M_{200c}^{\mathrm{match,sub}}$.

\begin{figure*}
\includegraphics[width=\columnwidth]{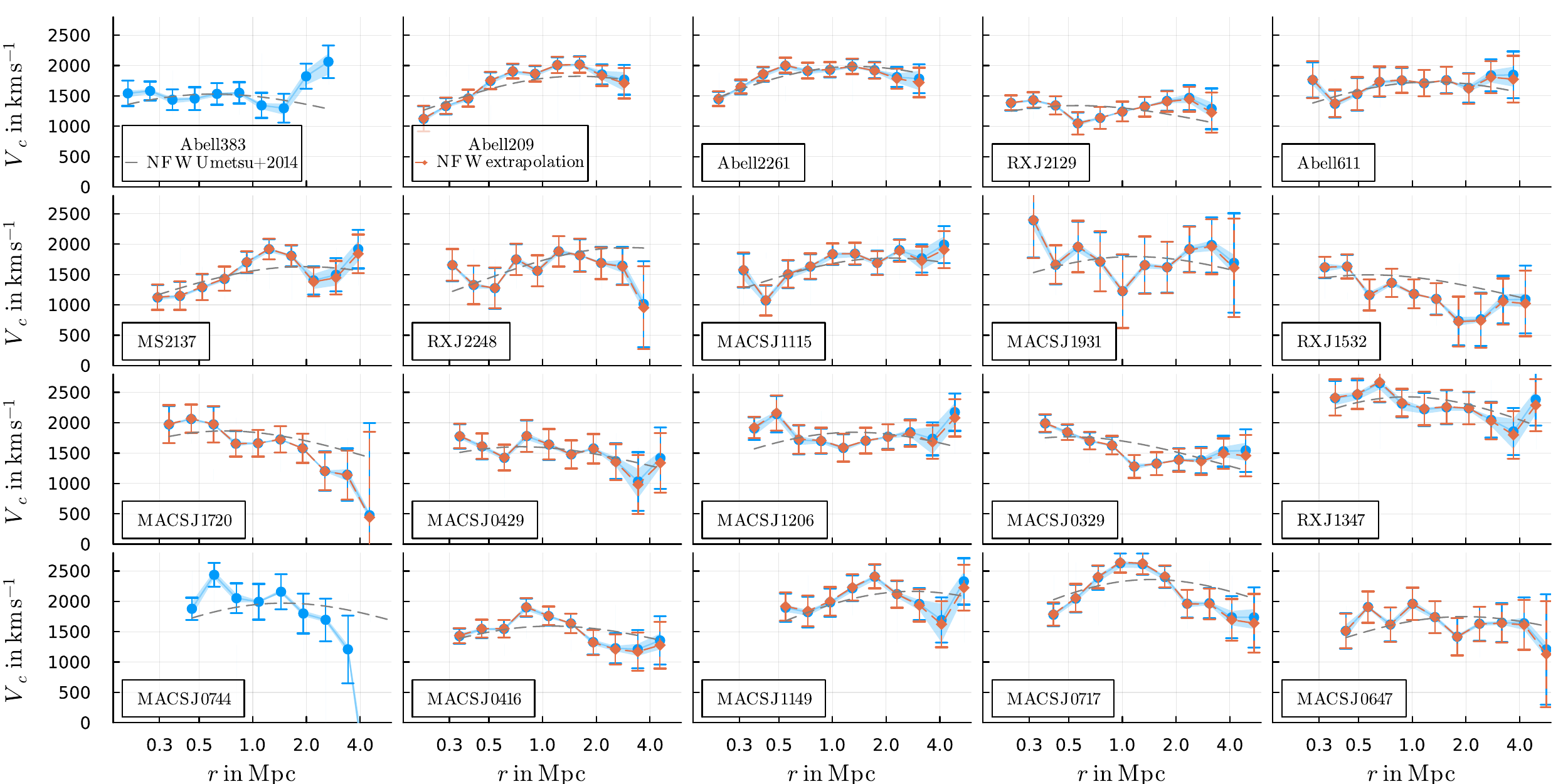}
\caption{%
 Same as Fig.~\ref{fig:vc-grid}, but additionally showing results when extrapolating the shear profile $G_+$ assuming an NFW profile (red symbols) instead of assuming a $1/R$ power law (blue symbols).
  We do not show results without the two-halo subtraction for visual clarity.
  Dashed gray lines are the circular velocities implied by the NFW fits from \citet{Umetsu2014}, which do not take into account the two-halo term.
  There is no result with NFW extrapolation for Abell 383 since no value for $M_{200c}$ can be determined from its non-parametric mass profile (see Sec.~\ref{sec:results:vc}) and we cannot switch to $M_{500c}$ because our mass-concentration relation works only for $M_{200c}$.
  Similarly, there is no result with NFW extrapolation for MACS J0744 because the last $G_+$ data point is negative, so our simple matching procedure does not find an NFW profile to extrapolate with.
 }
\label{fig:vc-nfw-grid}
\end{figure*}

Figure~\ref{fig:vc-nfw-grid} shows the cirular velocities inferred using the NFW extrapolation described above, including the adjusted two-halo subtraction procedure.
At large radii there is a small difference compared to our fiducial $1/R$ extrapolation.
This small difference is adequately captured by our systematic uncertainty band which is spanned by extrapolation with $1/\sqrt{R}$ and $1/R^2$ power laws.
For reference, Fig.~\ref{fig:vc-nfw-grid} also shows the circular velocities implied by the NFW fits from \citet{Umetsu2014}.
These NFW fits do not take into account the two-halo term.

\section{NFW fits to non-parametric mass and density profiles}
\label{sec:appendix:nfwfit}

\begin{figure}
\begin{center}
\includegraphics[width=.55\columnwidth]{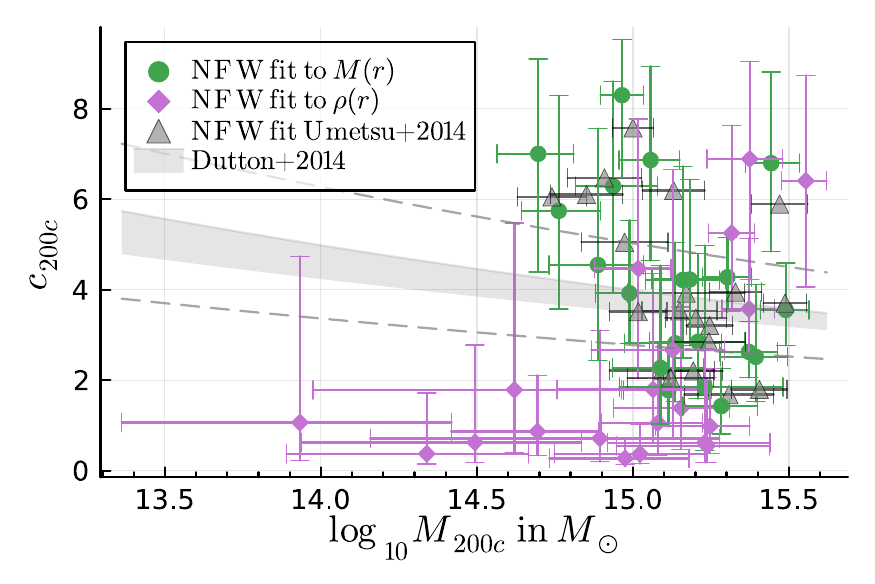}
\end{center}
\caption{%
 The best-fit parameters obtained by fitting NFW profiles to our non-parametric mass profiles $M(r)$ from Fig.~\ref{fig:M-grid} (green circles) and to our non-parametric density profiles $\rho(r)$ from Fig.~\ref{fig:rho-grid} (purple diamonds).
 Unlike the mass profiles $M(r)$, the density profiles $\rho(r)$ do not contain information about the mass distribution within the smallest radial bin, leading to different fit results.
 No two-halo subtraction is performed, allowing for a direct comparison to the NFW fits from \citet{Umetsu2014} (gray triangles).
 No uncertainties on the concentrations are given in \citet{Umetsu2014} so these are not shown.
 The shaded gray region indicates the mass-concentration relation from \citet{Dutton2014} for the range of redshifts of our cluster sample.
 Gray dashed lines indicate $0.1\,\mathrm{dex}$ scatter around that region.
}
\label{fig:c-M-NFW-fits}
\end{figure}

\begin{deluxetable}{l|cc|cc}
\tablecaption{NFW fit results}
\tablehead{
\colhead{} &
\multicolumn{2}{c}{NFW fit to $M(r)$} &
\multicolumn{2}{c}{NFW fit to $\rho(r)$}
\\
\colhead{Name} &
\colhead{$\log_{10} M_{200c}$} &
\colhead{$c_{200c}$} &
\colhead{$\log_{10} M_{200c}$} &
\colhead{$c_{200c}$}
\\
\colhead{} &
\colhead{$M_\odot$} &
\colhead{} &
\colhead{$M_\odot$} &
\colhead{}
}
\startdata
\input{./plots/table-nfw-fits.tex}\enddata
\tablecomments{%
 The listed values are the 16th, 50th, and 84th percentiles.
}
\label{tab:nfwfits}
\end{deluxetable}

We have fit NFW profiles to both our non-parametric mass profiles $M(r)$ and our non-parametric density profiles $\rho(r)$.
We loosely follow \citet{Umetsu2014} in using a Bayesian fitting procedure with flat priors for $13 < \log_{10} M_{200c}/M_\odot < 17$ and $-1 < \log_{10} c_{200c} < 1$.
We use the julia package `Turing.jl' \citep{Ge2018} with the `Emcee()' sampler \citep{Foreman-Mackey2013}.
For $\rho(r)$, but not $M(r)$, it can happen that the covariance matrix has a zero eigenvalue.
For example, a direct calculation shows that, in a simple setup with just two radial bins $R_1$ and $R_2$ and linear interpolation, it can happen that the inferred density in the first bin, $\rho(r=R_1)$, is completely independent of the shear measurement at $R_1$, due to a cancellation of different terms in Eq.~\eqref{eq:densityreconstruction}.
The reconstructed $\rho$ values in the two radial bins are then 100\% correlated; both are completely determined by the shear at $R_2$.
This leads to a zero eigenvalue in the covariance matrix.
To avoid numerical issues with the inverse covariance matrix, we detect such behavior and remove the corresponding eigenvector from the fit, leaving only the orthogonal subspace of densities to be fit.
We apply this removal procedure when the smallest eigenvalue of the correlation matrix is at least $100$ times smaller than the second-smallest eigenvalue.

The best-fit parameters are shown in Fig.~\ref{fig:c-M-NFW-fits} and Table~\ref{tab:nfwfits}.
We did not apply our two-halo subtraction procedure to allow a more direct comparison to \citet{Umetsu2014}.
When fitting to $M(r)$, we find best-fit parameters consistent with those of \citet{Umetsu2014}.

However, fitting our non-parametric \emph{density} profiles does not recover the same fit parameters.
In particular, concentrations are systematically smaller when fitting $\rho(r)$.
In addition, the statistical uncertainties on the NFW parameters are larger, with concentrations being particularly poorly constrained.
This is because in going from $M(r)$ to $\rho(r)$ one loses information, unless $\rho(r)$ is measured all the way down to $r=0$:
Weak-lensing observations do not extend all the way to the centers of clusters, so there is a minimum radius $r_{\mathrm{min}}$ down to which we infer $\rho(r)$.
As a result, we cannot reconstruct $M(r)$ from $\rho(r)$.
We can only reconstruct $M(r) - M(r_{\mathrm{min}})$, because the density $\rho(r)$ at $r > r_{\mathrm{min}}$ does not know anything about the mass distribution within $r_{\mathrm{min}}$.
In contrast, the mass profile $M(r)$ at $r > r_{\mathrm{min}}$ does.
It knows the total amount of mass within $r_{\mathrm{min}}$.

There is some amount of degeneracy between concentration and mass even when fitting to $M(r)$.
This may be why the \citet{Dutton2014} mass-concentration relation shown in Fig.~\ref{fig:c-M-NFW-fits} seems not to be followed very closely even by our fits to $M(r)$.
Breaking this degeneracy may require extending the radial range covered by observations, for example using strong lensing \citep[e.g.,][]{Merten2015,Umetsu2016,Umetsu2025}.

\section{Missing mass assuming the galaxy-scale RAR}
\label{sec:appendix:missing}

\begin{figure*}
\begin{center}
\includegraphics[width=.8\columnwidth]{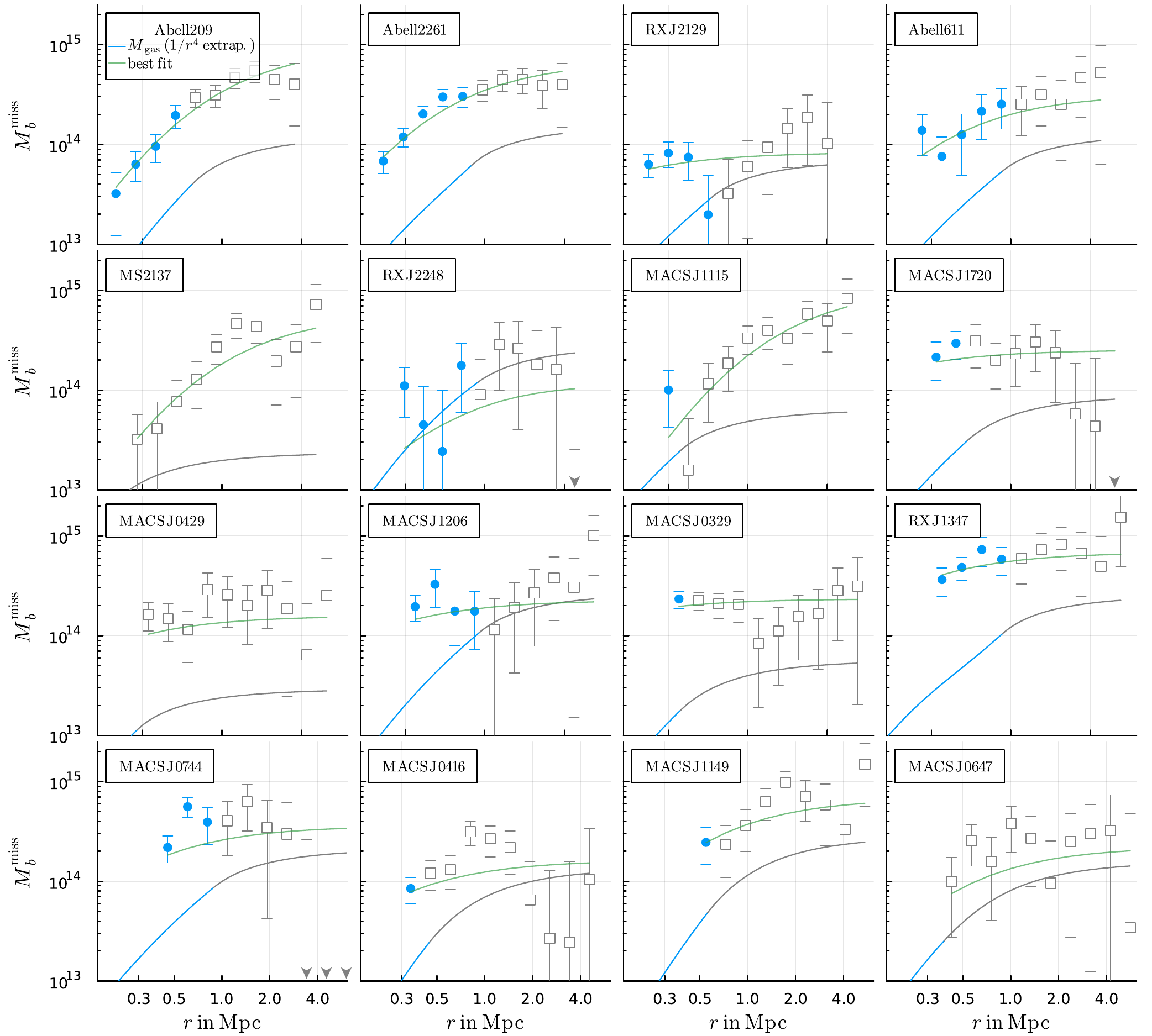}
\end{center}
\caption{%
 The missing baryonic mass implied by assuming that the galaxy-scale RAR holds also for galaxy clusters.
 Blue circles indicate the radial range where X-ray and weak-lensing observations overlap.
 Beyond that range, we extrapolate the gas densities assuming a $1/r^4$ tail (gray squares).
 Arrowheads at the horizontal axis indicate negative missing mass.
 We do not subtract the two-halo term.
 Solid blue and gray lines show the gas mass profiles.
 The solid green line indicates the best fit to $M_b^{\mathrm{miss}}$ following the fitting procedure of \citet{Kelleher2024}.
}
\label{fig:missing-Mb-quartic}
\end{figure*}

The galaxy-scale RAR can be parametrized as a relation between $g_{\mathrm{obs}}$ and $g_{\mathrm{bar}}$ with $\mu(|g_{\mathrm{obs}}|/a_0) \, g_{\mathrm{obs}} = g_{\mathrm{bar}}$, where $\mu$ is the so-called interpolation function.
We adopt the so-called ``simple'' interpolation function $\mu(x) = x/(1+x)$.
If we assume the galaxy-scale RAR to hold universally, the missing baryonic mass $M_b^{\mathrm{miss}}$ is then given by
\begin{equation}
 \frac{G_N M_b^{\mathrm{miss}}(r)}{r^2} =\mu\left(\frac{|g_{\mathrm{obs}} (r)|}{a_0}\right) g_{\mathrm{obs}} (r) - g_{\mathrm{bar}} (r) \,.
\end{equation}
This $M_b^{\mathrm{miss}}$ must be a monotonic function of radius if the mismatch between our measurements of $\mu(|g_{\mathrm{obs}}|) g_{\mathrm{obs}}$ and $g_{\mathrm{bar}}$ is indeed due to missing baryons.

\begin{figure*}
\begin{center}
\includegraphics[width=.8\columnwidth]{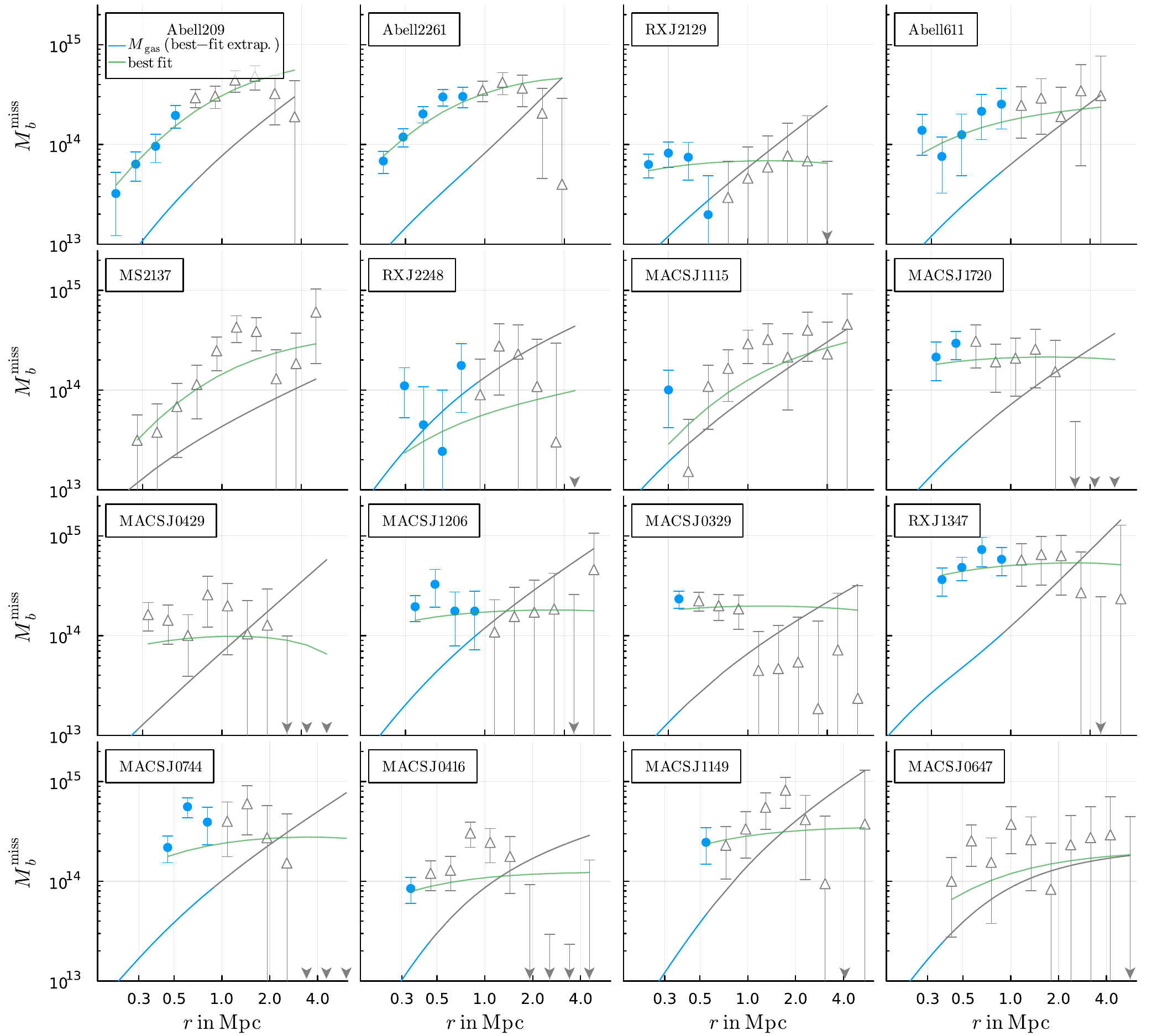}
\end{center}
\caption{%
 Same as Fig.~\ref{fig:missing-Mb-quartic} but extrapolating the gas densities assuming the best-fit beta profiles.
}
\label{fig:missing-Mb-beta}
\end{figure*}

Figure~\ref{fig:missing-Mb-quartic} shows the missing mass $M_b^{\mathrm{miss}}$ implied by our weak-lensing analysis and assuming a $1/r^4$ extrapolation for the gas densities.
We do not subtract the two-halo term.
Subtracting the two-halo term does not significantly change the results.
In the region where X-ray and weak-lensing observations overlap, we find that $M_b^{\mathrm{miss}}$ is a monotonic function within the uncertainties.
An exception are the innermost three data points of RX J2129, which, however, may not be reliable (Sec.~\ref{sec:results:vc}).
At larger radii, some clusters have non-monotonic and even negative $M_b^{\mathrm{miss}}$, but the error bars are large and often consistent with the monotonic mass profiles $M_{\mathrm{mm}}(r)$ (solid green lines, see below).

Figure~\ref{fig:missing-Mb-beta} shows the $M_b^{\mathrm{miss}}$ implied by assuming the beta profile fits from \citet{Famaey2024} are valid even beyond $R_{\mathrm{max}}^X$.
This increases $g_{\mathrm{bar}}$ at large radii, implying more non-monotonicities in $M_b^{\mathrm{miss}}$.
\citet{Famaey2024} find that, within $1\,\mathrm{Mpc}$, $M_b^{\mathrm{miss}}$ tracks the gas mass.
We can confirm that result for a few clusters, for example for Abell 209 and Abell 2261, but it does not seem to hold as univerally as for the NFW fits from \citet{Famaey2024}.

The solid green lines in Fig.~\ref{fig:missing-Mb-quartic} and Fig.~\ref{fig:missing-Mb-beta} show the best fit missing mass component $M_{\mathrm{mm}}(r)$ following \citet{Kelleher2024} (Sec.~\ref{sec:results:missing}).
This additional mass component $M_{\mathrm{mm}}(r)$ is always a monotonic function of $r$.
Despite this, the solid green lines in Fig.~\ref{fig:missing-Mb-beta} can be non-monotonic because they correspond to $M_{\mathrm{mm}}(r) + (\Upsilon_b - 1) M_b(r)$ which is the quantity that, for good fits, should match the non-parametric missing mass profiles shown there (which have $\Upsilon_b=1$).

\end{appendix}

\end{document}

%% file: plots/table-cluster-list.tex
Abell 383 & $0.187$ & $-$  & $-$  & $-$ \\ 
Abell 209 & $0.206$ & $0.651$ & $15.26 \pm 0.11$ & $15.29 \pm 0.12$\\ 
Abell 2261 & $0.224$ & $0.825$ & $15.25 \pm 0.11$ & $15.28 \pm 0.12$\\ 
RXJ2129 & $0.234$ & $0.591^\ast$ & $14.93 \pm 0.18$ & $14.96 \pm 0.18$\\ 
Abell 611 & $0.288$ & $0.884$ & $15.11 \pm 0.15$ & $15.15 \pm 0.38$\\ 
MS2137 & $0.313$ & $0.227$ & $15.05 \pm 0.08$ & $15.07 \pm 0.08$\\ 
RXJ2248 & $0.348$ & $0.904$ & $15.14 \pm 0.15$ & $15.17 \pm 0.16$\\ 
MACSJ1115 & $0.352$ & $0.375$ & $15.27 \pm 0.18$ & $15.32 \pm 0.18$\\ 
MACSJ1931 & $0.352$ & $-$  & $15.27 \pm 0.46$ & $15.33 \pm 0.31$\\ 
RXJ1532 & $0.363$ & $-$  & $14.58 \pm 0.24$ & $14.60 \pm 0.25$\\ 
MACSJ1720 & $0.391$ & $0.532$ & $15.04 \pm 0.15$ & $15.07 \pm 0.11$\\ 
MACSJ0429 & $0.399$ & $0.292$ & $15.02 \pm 0.23$ & $15.06 \pm 0.24$\\ 
MACSJ1206 & $0.440$ & $0.930$ & $15.17 \pm 0.17$ & $15.22 \pm 0.17$\\ 
MACSJ0329 & $0.450$ & $0.377$ & $14.79 \pm 0.20$ & $14.83 \pm 0.23$\\ 
RXJ1347 & $0.451$ & $0.887$ & $15.41 \pm 0.11$ & $15.44 \pm 0.12$\\ 
MACSJ0744 & $0.686$ & $0.884$ & $15.17 \pm 0.13$ & $15.19 \pm 0.14$\\ 
MACSJ0416 & $0.396$ & $0.449$ & $14.92 \pm 0.09$ & $14.94 \pm 0.09$\\ 
MACSJ1149 & $0.544$ & $0.557$ & $15.38 \pm 0.10$ & $15.42 \pm 0.10$\\ 
MACSJ0717 & $0.548$ & $-$  & $15.32 \pm 0.08$ & $15.35 \pm 0.08$\\ 
MACSJ0647 & $0.584$ & $0.381$ & $14.94 \pm 0.13$ & $14.97 \pm 0.13$ 

%% file: plots/table-missing-Mb-fits.tex
Abell 209 & $14.99_{-0.16}^{+0.15}$ & $15.48_{-0.25}^{+0.25}$ & $2.63_{-0.13}^{+0.12}$ & $+0.01_{-0.10}^{+0.10}$ & $14.91_{-0.18}^{+0.16}$ & $15.56_{-0.26}^{+0.27}$ & $2.58_{-0.14}^{+0.13}$ & $-0.01_{-0.10}^{+0.10}$\\ 
Abell 2261 & $14.83_{-0.14}^{+0.13}$ & $16.02_{-0.22}^{+0.24}$ & $2.40_{-0.12}^{+0.11}$ & $+0.00_{-0.10}^{+0.10}$ & $14.79_{-0.15}^{+0.13}$ & $16.09_{-0.22}^{+0.25}$ & $2.36_{-0.12}^{+0.11}$ & $-0.03_{-0.09}^{+0.09}$\\ 
RXJ2129 & $13.92_{-0.16}^{+0.17}$ & $17.75_{-1.08}^{+0.94}$ & $1.50_{-0.34}^{+0.41}$ & $+0.00_{-0.10}^{+0.10}$ & $13.88_{-0.15}^{+0.15}$ & $17.89_{-1.03}^{+0.87}$ & $1.45_{-0.31}^{+0.38}$ & $-0.02_{-0.09}^{+0.09}$\\ 
Abell 611 & $14.49_{-0.33}^{+0.36}$ & $16.23_{-0.91}^{+1.58}$ & $2.21_{-0.64}^{+0.41}$ & $+0.02_{-0.10}^{+0.10}$ & $14.39_{-0.29}^{+0.37}$ & $16.48_{-1.03}^{+1.67}$ & $2.09_{-0.65}^{+0.45}$ & $+0.02_{-0.10}^{+0.10}$\\ 
MS2137 & $14.77_{-0.30}^{+0.29}$ & $15.13_{-0.55}^{+0.57}$ & $2.68_{-0.28}^{+0.26}$ & $+0.00_{-0.10}^{+0.10}$ & $14.57_{-0.48}^{+0.36}$ & $15.22_{-0.74}^{+0.92}$ & $2.58_{-0.45}^{+0.33}$ & $+0.02_{-0.11}^{+0.11}$\\ 
RXJ2248 & $13.92_{-1.08}^{+0.57}$ & $15.58_{-3.14}^{+2.33}$ & $2.23_{-0.85}^{+1.04}$ & $+0.05_{-0.10}^{+0.10}$ & $13.84_{-1.08}^{+0.57}$ & $15.58_{-3.33}^{+2.26}$ & $2.19_{-0.81}^{+1.13}$ & $+0.03_{-0.10}^{+0.09}$\\ 
MACSJ1115 & $15.03_{-0.28}^{+0.27}$ & $14.89_{-0.51}^{+0.50}$ & $2.84_{-0.24}^{+0.24}$ & $+0.01_{-0.10}^{+0.10}$ & $14.52_{-0.84}^{+0.49}$ & $15.07_{-1.14}^{+1.35}$ & $2.63_{-0.69}^{+0.45}$ & $+0.05_{-0.11}^{+0.11}$\\ 
MACSJ1720 & $14.40_{-0.15}^{+0.14}$ & $18.19_{-1.25}^{+1.04}$ & $1.52_{-0.36}^{+0.45}$ & $+0.00_{-0.10}^{+0.10}$ & $14.38_{-0.16}^{+0.14}$ & $18.23_{-1.21}^{+0.99}$ & $1.49_{-0.34}^{+0.43}$ & $-0.03_{-0.10}^{+0.10}$\\ 
MACSJ0429 & $14.20_{-0.22}^{+0.22}$ & $17.39_{-1.46}^{+1.37}$ & $1.71_{-0.48}^{+0.55}$ & $+0.01_{-0.10}^{+0.10}$ & $14.07_{-0.32}^{+0.21}$ & $17.56_{-1.67}^{+1.23}$ & $1.60_{-0.42}^{+0.60}$ & $-0.04_{-0.09}^{+0.10}$\\ 
MACSJ1206 & $14.34_{-0.19}^{+0.22}$ & $17.51_{-1.22}^{+1.31}$ & $1.72_{-0.47}^{+0.47}$ & $+0.01_{-0.10}^{+0.10}$ & $14.29_{-0.19}^{+0.19}$ & $17.79_{-1.28}^{+1.20}$ & $1.61_{-0.42}^{+0.49}$ & $-0.01_{-0.09}^{+0.09}$\\ 
MACSJ0329 & $14.37_{-0.09}^{+0.08}$ & $18.69_{-0.98}^{+0.71}$ & $1.35_{-0.25}^{+0.34}$ & $-0.01_{-0.10}^{+0.10}$ & $14.34_{-0.09}^{+0.08}$ & $18.76_{-0.88}^{+0.65}$ & $1.31_{-0.22}^{+0.31}$ & $-0.04_{-0.09}^{+0.09}$\\ 
RXJ1347 & $14.83_{-0.13}^{+0.19}$ & $17.71_{-1.01}^{+1.45}$ & $1.83_{-0.53}^{+0.39}$ & $+0.01_{-0.10}^{+0.10}$ & $14.77_{-0.12}^{+0.16}$ & $18.06_{-1.09}^{+1.32}$ & $1.69_{-0.47}^{+0.42}$ & $-0.02_{-0.10}^{+0.09}$\\ 
MACSJ0744 & $14.55_{-0.24}^{+0.28}$ & $16.78_{-0.94}^{+1.66}$ & $2.05_{-0.63}^{+0.40}$ & $+0.01_{-0.10}^{+0.10}$ & $14.51_{-0.24}^{+0.28}$ & $16.86_{-1.01}^{+1.60}$ & $2.00_{-0.61}^{+0.42}$ & $-0.02_{-0.10}^{+0.10}$\\ 
MACSJ0416 & $14.19_{-0.22}^{+0.26}$ & $16.72_{-0.88}^{+1.45}$ & $1.94_{-0.55}^{+0.37}$ & $+0.02_{-0.10}^{+0.10}$ & $14.12_{-0.20}^{+0.25}$ & $17.07_{-1.05}^{+1.37}$ & $1.80_{-0.50}^{+0.43}$ & $-0.01_{-0.09}^{+0.09}$\\ 
MACSJ1149 & $14.82_{-0.29}^{+0.31}$ & $16.19_{-0.92}^{+1.88}$ & $2.34_{-0.73}^{+0.40}$ & $+0.02_{-0.10}^{+0.10}$ & $14.56_{-0.25}^{+0.28}$ & $17.10_{-1.45}^{+1.75}$ & $1.92_{-0.62}^{+0.58}$ & $-0.00_{-0.09}^{+0.09}$\\ 
MACSJ0647 & $14.32_{-0.66}^{+0.47}$ & $15.90_{-1.87}^{+1.97}$ & $2.25_{-0.78}^{+0.71}$ & $+0.03_{-0.10}^{+0.11}$ & $14.26_{-0.75}^{+0.49}$ & $15.83_{-2.18}^{+2.02}$ & $2.26_{-0.79}^{+0.77}$ & $+0.04_{-0.10}^{+0.10}$ 

%% file: plots/table-nfw-fits.tex
Abell 383 & $14.94_{-0.12}^{+0.14}$ & $6.29_{-2.34}^{+2.32}$ & $15.02_{-0.27}^{+0.21}$ & $0.37_{-0.22}^{+0.66}$\\ 
Abell 209 & $15.37_{-0.09}^{+0.09}$ & $2.63_{-0.57}^{+0.67}$ & $15.37_{-0.09}^{+0.08}$ & $3.58_{-1.07}^{+1.56}$\\ 
Abell 2261 & $15.30_{-0.08}^{+0.08}$ & $4.28_{-0.76}^{+0.88}$ & $15.32_{-0.07}^{+0.07}$ & $5.25_{-1.72}^{+2.38}$\\ 
RXJ2129 & $14.76_{-0.12}^{+0.13}$ & $5.74_{-2.17}^{+2.55}$ & $14.69_{-0.28}^{+0.20}$ & $0.87_{-0.53}^{+1.24}$\\ 
Abell 611 & $15.21_{-0.15}^{+0.15}$ & $2.85_{-1.26}^{+1.94}$ & $15.15_{-0.22}^{+0.16}$ & $1.38_{-0.91}^{+2.23}$\\ 
MS2137 & $15.11_{-0.14}^{+0.13}$ & $1.78_{-0.80}^{+1.07}$ & $15.08_{-0.18}^{+0.14}$ & $1.05_{-0.71}^{+1.75}$\\ 
RXJ2248 & $15.14_{-0.16}^{+0.15}$ & $2.82_{-1.29}^{+2.23}$ & $15.06_{-0.31}^{+0.19}$ & $1.79_{-1.17}^{+2.56}$\\ 
MACSJ1115 & $15.28_{-0.12}^{+0.12}$ & $1.43_{-0.62}^{+0.84}$ & $15.25_{-0.17}^{+0.13}$ & $0.98_{-0.60}^{+1.27}$\\ 
MACSJ1931 & $15.23_{-0.27}^{+0.25}$ & $1.85_{-1.40}^{+3.13}$ & $14.89_{-0.74}^{+0.38}$ & $0.71_{-0.52}^{+2.39}$\\ 
RXJ1532 & $14.70_{-0.13}^{+0.11}$ & $7.01_{-2.62}^{+2.09}$ & $13.93_{-0.57}^{+0.49}$ & $1.06_{-0.84}^{+3.68}$\\ 
MACSJ1720 & $15.06_{-0.10}^{+0.09}$ & $6.87_{-2.22}^{+2.07}$ & $14.62_{-0.65}^{+0.34}$ & $1.79_{-1.39}^{+3.69}$\\ 
MACSJ0429 & $14.89_{-0.16}^{+0.15}$ & $4.55_{-2.11}^{+3.02}$ & $14.49_{-0.56}^{+0.34}$ & $0.62_{-0.44}^{+2.16}$\\ 
MACSJ1206 & $15.16_{-0.12}^{+0.12}$ & $4.22_{-1.73}^{+2.51}$ & $14.97_{-0.24}^{+0.20}$ & $0.27_{-0.13}^{+0.43}$\\ 
MACSJ0329 & $14.96_{-0.07}^{+0.07}$ & $8.30_{-1.82}^{+1.23}$ & $14.34_{-0.45}^{+0.33}$ & $0.37_{-0.22}^{+1.35}$\\ 
RXJ1347 & $15.44_{-0.08}^{+0.09}$ & $6.80_{-1.95}^{+2.01}$ & $15.23_{-0.31}^{+0.21}$ & $0.62_{-0.43}^{+2.08}$\\ 
MACSJ0744 & $15.18_{-0.12}^{+0.11}$ & $4.23_{-1.46}^{+2.21}$ & $15.37_{-0.14}^{+0.11}$ & $6.89_{-2.67}^{+2.15}$\\ 
MACSJ0416 & $14.99_{-0.11}^{+0.10}$ & $3.92_{-1.12}^{+1.61}$ & $15.02_{-0.14}^{+0.11}$ & $4.47_{-2.43}^{+3.31}$\\ 
MACSJ1149 & $15.39_{-0.11}^{+0.10}$ & $2.52_{-0.99}^{+1.60}$ & $15.24_{-0.29}^{+0.20}$ & $0.55_{-0.36}^{+1.50}$\\ 
MACSJ0717 & $15.49_{-0.08}^{+0.07}$ & $3.55_{-0.79}^{+1.04}$ & $15.55_{-0.08}^{+0.07}$ & $6.41_{-2.35}^{+2.33}$\\ 
MACSJ0647 & $15.09_{-0.15}^{+0.14}$ & $2.27_{-1.24}^{+2.26}$ & $15.13_{-0.26}^{+0.17}$ & $2.67_{-1.94}^{+3.99}$ 